\documentclass{aa}  

\usepackage{graphicx}
\usepackage{txfonts}
\usepackage{multirow}
\usepackage{multicol}
\newcommand{\hii}{H\textsc{ii}}
\def\ks{km s$^{-1}$}

\def\cm3{cm$^{-3}$}

\def\2{$^{12}$CO}
\def\3{$^{13}$CO}
\def\msol{M$_\odot$}

\begin{document}

   \title{$^{12}$CO and $^{13}$CO J=3--2 observations toward N11 in the Large Magellanic Cloud}

\author{M. Celis Peña\inst{1}
          \and
          S. Paron\inst{1}
          \and
          M. Rubio\inst{2}
          \and
          C. N. Herrera\inst{3}
          \and
          M. E. Ortega\inst{1}
          }

\institute{CONICET - Universidad de Buenos Aires, Instituto de Astronom\'{\i}a y  F\'{\i}sica del Espacio, CC 67, Suc. 28, 1428 Buenos Aires, Argentina\\
              \email{mcelis@iafe.uba.ar}
         \and
             Departamento de Astronom\'ia, Universidad de Chile, Casilla 36-D, Santiago, Chile
         \and
             Institut de Radioastronomie Millimétrique, 300 Rue de la Piscine, 38406 Saint-Martin-d'Hères, France
             }

   \date{Received  <date>; Accepted <date> }

\abstract{}
{After 30 Doradus, N11 is the second largest and brightest nebula in the Large Magellanic
Cloud (LMC). This large nebula has several OB associations with bright nebulae at its surroundings. 
N11 was previously mapped at the lowest rotational transitions of $^{12}$CO (J=1--0 and 2--1), and in some particular
regions pointings of the $^{13}$CO J=1--0 and 2--1 lines were also performed. 
Observations of higher CO rotational transitions are needed to map gas with higher critical densities, useful to study in a more
accurate way the physical conditions of the gas component and its relation with the UV radiation.}
{Using the Atacama Submillimeter Telescope Experiment we mapped the whole extension of the N11 nebula in the $^{12}$CO J=3--2 line,
and three sub-regions in the $^{13}$CO J=3--2 line. 
The regions mapped in the $^{13}$CO J=3--2 were selected based on that they may be exposed 
to the radiation at different ways: a region lying over the nebula related to the OB association LH10 (N11B), 
another one that it is associated with the southern part of the nebula related to the OB association LH13 (N11D), and finally a farther 
area at the southwest without any embedded OB association (N11I).}
{We found that the morphology of the molecular clouds lying in each region shows some
signatures that could be explained by  the expansion
of the nebulae and the action of the radiation. Fragmentation generated in a molecular shell due to
the expansion of the N11 nebula is suggested. The integrated line ratios \2/\3 show evidences of selective
photodissociation of the \3, and probably other mechanisms such as chemical fractionation. The values found for the integrated line ratios 
\2 J=3--2/1--0 are in agreement with values that were assumed in previous works, and the CO contribution to the continuum 
at 870 $\mu$m was directly derived. The distribution of the integrated line ratios \2 J=3--2/2--1 show hints of stellar feedback in N11B and N11D. 
The ratio between the virial and LTE mass (M$_{\rm vir}$/M$_{\rm LTE}$) is higher 
than unity in all analyzed molecular clumps, which suggests that the clumps are not gravitationally bounded and may be supported by external 
pressure. A non-LTE analysis suggests that we are mapping gas with densities about a few 10$^{3}$ cm$^{-3}$. The molecular 
clump at N11B, the unique molecular feature with direct evidence of ongoing star formation, is the densest one among the analyzed clumps.
}
{}

\titlerunning{$^{12}$CO and $^{13}$CO J=3--2 observations toward N11}
\authorrunning{M. Celis Peña et al.}

\keywords{galaxies: ISM -- Magellanic Clouds -- HII regions -- ISM: individual objects: N11}

   \maketitle
%

\section{Introduction}

Studying low-metallicity interstellar media toward other galaxies, whose physical 
conditions may resemble those that existed in the early Milky Way, is a very important 
issue because it can shed light on the primeval processes of star formation that occurred 
in our Galaxy.

Due to its proximity (about 50 kpc) the Large Magellanic Cloud (LMC) is 
an ideal laboratory to study the physical properties of molecular clouds under
conditions different from those found in our Galaxy. 
Indeed, Magellanic Clouds are a unique laboratory to study the effects of 
metallicity and galaxy mass on molecular gas and star formation at high spatial
resolution \citep{jameson2016}.
The metallicity in the LMC is Z $\sim$ 0.5 Z$_{\odot}$ \citep{keller2006} and the gas-to-dust ratio is a factor of 4 higher than 
in our Galaxy. As the LMC is seen nearly face-on with an inclination angle of 35$^{\circ}$, many active star-forming and \hii~regions
were found and studied (e.g. \citealt{ochsendorf2017}, \citealt{paron2015,paron2014}).

\begin{figure}[h]
  \centering
  \includegraphics[width=8.7cm]{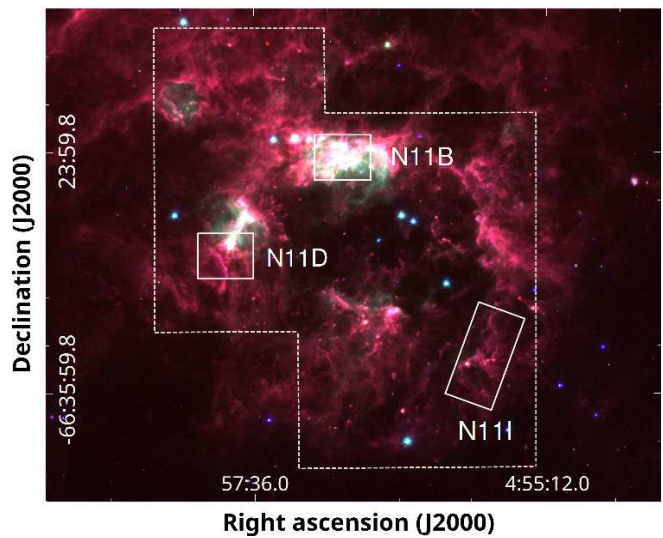}
  \caption{Three-colour image where the 3.6, 4.5, and 8 $\mu$m emission obtained from the IRAC camera of the {\it Spitzer} Space Telescope
(from SAGE Spitzer) are presented in blue, green and red, respectively. The white boxes show the regions mapped in the \3 J=3--2 line,
while the area delimited by the dashed line is the region surveyed in the \2 J=3--2 line. }
  \label{im_intro}
\end{figure}

After 30 Doradus, N11 \citep{henize56} is the second largest and brightest nebula of the LMC. It is located at the north-western corner and it is
one of the most important star-forming region in the galaxy. 
N11 has a ring morphology (see Fig.\,\ref{im_intro}) with a cavity of 170 pc in diameter enclosing the OB association LH9 \citep{lucke1970} 
also known as NGC 1760. This region also presents several OB associations with bright nebulae on its surroundings \citep{rosado96}; 
for instance, LH10 (NGC 1763) lying at the northern rim and exciting the N11B nebula, and LH13 (NGC 1769) at the east and associated
with the N11D nebula. \citet{parker1992} and \citet{walb92} proposed sequential star formation among the OB associations;
LH9, an older association, may have generated the formation of LH10, which in turn is likely triggering new star formation in the 
N11B nebula surroundings \citep{barba03}. Moreover, \citet{hatano06} have proposed that LH9 is indeed triggering star 
formation in the surroundings molecular clouds.
Concerning the molecular gas related to N11, \citet{israel03} and \citet{herrera2013} provided catalogues of physical properties of 
individual molecular clouds distributed through the N11 complex based on $^{12}$CO J=1--0 and J=2--1 data.
They pointed out that the molecular gas component related to N11 exhibits a shell morphology with massive clumps embedded, 
which is the usual configuration for a triggered star formation scenario around \hii~regions (e.g. \citealt{elmegreen77,pomares09}).
Additionally, \citet{israel11} observed that the distribution of CO and [CII] emissions are quite similar, suggesting that a large-scale 
dissociation of CO and subsequent ionization of the resulting neutral carbon is indeed ongoing in the region.
More recently,  using {\it Spitzer}, {\it Herschel} and APEX/LABOCA data, \citet{galametz16} performed a complete study about the dust properties
toward N11 and investigated variations of the gas-to-dust ratio across the region. 

Observations of other CO isotopes and higher rotational transitions of $^{12}$CO are needed to map gas with 
lower optical depths and higher critical densities, useful to study, in a more
accurate way, the physical conditions of the gas component. 
In this paper, we present for the first time \2 and \3 J=3--2 observations toward N11. The \2 J=3--2 line, tracer of warmer and denser gas 
than \2 J=1--0, was observed in the entire N11 region (see the region enclosed by dashed lines in Fig.\,\ref{im_intro}). \3 J=3--2 observations 
were done toward three sub-regions  
(see boxes in Fig.\,\ref{im_intro}), which according to \citet{herrera2013}, have the highest CO peak 
temperatures and contain molecular clouds at different evolutionary stages.

\section{Presentation of the analyzed sub-regions in N11}

In this section we briefly describe the sub-regions mapped in the \3 J=3--2 line (see Fig.\,\ref{im_intro}).
 
The first one is a molecular cloud in the northern part of N11, N11B, associated with LH10, a young (3~Myr) OB association. LH10 is 
younger than LH9 \citep{walb99,mokiem07}, and \citet{parker1992} discover
that its stellar content presents a higher ratio of higher-mass to lower-mass stars than in LH9. 
This cloud has bright free-free emission, [CII] 158 $\mu$m and 8 $\mu$m emission. According to \citet{galametz16}, it shows the highest 
radiation field across N11.
\citet{israel11} obtain a $G_{0}$ value, which is the flux over the range
6--13.6 eV normalized to $1.6\times10^{-3}$ erg cm$^{-2}$ s$^{-1}$, of about 180, the highest in the region.
There is evidence of ongoing star formation in N11B. About 20 HAeBe stars, which are intermediate-mass
pre-main sequence stars, have been detected in this region \citep{barba03,hatano06}.
 
Second, N11D, a molecular cloud with bright free-free emission but not massive star formation (no direct evidence of star formation activity 
is observed in this region) in the eastern border of N11. This cloud is in the southern part of the nebula 
generated by the OB association LH13 \citep{rosado96}. \citet{galametz16} show that the 
region also has a significant radiation field intensity but not so extended nor intense than in the former described region.
\citet{israel11} obtain $G_{0}$ about 26 toward this region. 

Third, N11I, a molecular cloud in the south-western border of N11, that does not have bright free-free emission, 
indicating more quiescent molecular gas.

These clouds can be placed in the evolutionary sequence of molecular 
clouds proposed by \citet{kawamura09}: N11B would be a type III cloud (clouds associated with young stars), N11D, a type II 
(clouds associated with \hii~regions), and N11I, a type I (no \hii~region associated).

The center and sizes of the observed
regions in \3 J=3--2, and some of the above mentioned characteristics, are summarized in Table\,\ref{subregs}.

\section{ASTE observations and data reduction}

The observations of the $^{12}$CO $J$=3--2 emission line, mapping the entire N11 region (see the region enclosed by dashed lines in 
Fig.\,\ref{im_intro}), were performed during November and December 2014
with the 10 m Atacama Submillimeter Telescope Experiment (ASTE). The CATS345 GHz band receiver used is a two single-band SIS receiver 
remotely tunable in the LO frequency range of 324-372 GHz. The spectrometer MAC was used with a bandwidth of 128 MHz and a natural spectral 
resolution of 0.125 MHz. The spectral velocity resolution was 0.11 km s$^{-1}$ and the half-power beamwidth (HPBW)
was 22\arcsec at 345 GHz. We mapped the N11 region in the on-the-fly mapping mode covering the region with two different maps in both
X and Y scan directions in order to remove any residual on the maps due to the scan pattern.
The system temperatures values changed from 300 to 700 K during the observations. Pointing was performed toward R Dor, every 1 or 1.5 hours, 
depending on the weather conditions. We also observe Orion KL to check the absolute value of the calibrations, which agrees within 10\%.

During August 2015, three regions toward N11 (boxes in Fig.\,\ref{im_intro}, and see Table\,\ref{subregs}) were mapped in the $^{13}$CO J=3--2 
line using the same telescope 
with the CATS345 receiver and the MAC spectrometer with the same configuration as explained above. The spectral velocity resolution was 0.11 km s$^{-1}$ 
and the HPBW was 23\arcsec at 330 GHz. The observations were done also in the on-the-fly 
mapping mode. The system temperature was between 200 and 400 K. In both cases the main beam efficiency was about $\eta_{mb}=$ 0.6.
 
The observations were reduced with NOSTAR \citep{sawada2008} and some spectra processed using the XSpec software
 package\footnote{XSpec is a spectral 
line reduction package for astronomy which has been developed by Per Bergman at Onsala Space Observatory.}. The spectra were Hanning-smoothed 
to improve the signal-to-noise ratio and low-degree polynomials were used for baseline fitting. The typical rms noise level is 200 and 40 mK
for the \2 and \3, respectively.

\begin{table*}
\caption{Observed regions in $^{13}$CO J=3--2 toward N11.}
\label{subregs} 
\centering
\begin{tabular}{l c c c l c}
\hline\hline 
   Region	&  RA (J2000)   & Dec.(J2000) &	Size  & OB association  & MCs evolutionary sequence\tablefootmark{\bf a} \\
\hline                        
N11B & 04$^\textrm{h}$ 56$^\textrm{m}$ 48$^\textrm{s}$ & -66$^\circ$ 24$^\prime$ 25$^{\prime\prime}$ & 140$^{\prime\prime}\times$110$^{\prime\prime}$		&  LH10 (embedded)  & type III \\
N11D & 04$^\textrm{h}$ 55$^\textrm{m}$ 41$^\textrm{s}$ & -66$^\circ$ 34$^\prime$ 25$^{\prime\prime}$ & 130$^{\prime\prime}\times$100$^{\prime\prime}$		&  LH13 (to the north) & type II \\
N11I & 04$^\textrm{h}$ 57$^\textrm{m}$ 50$^\textrm{s}$ & -66$^\circ$ 28$^\prime$ 57$^{\prime\prime}$ & 80$^{\prime\prime}\times$230$^{\prime\prime}$		& none & type I \\
\hline
\end{tabular}
\tablefoot{
\tablefoottext{\bf a}{probable evolutionary sequence (from \citealt{kawamura09}) of the molecular clouds in the regions.}}
\end{table*}

\section{Results}

\begin{figure}[h]
  \centering
  \includegraphics[width=9cm]{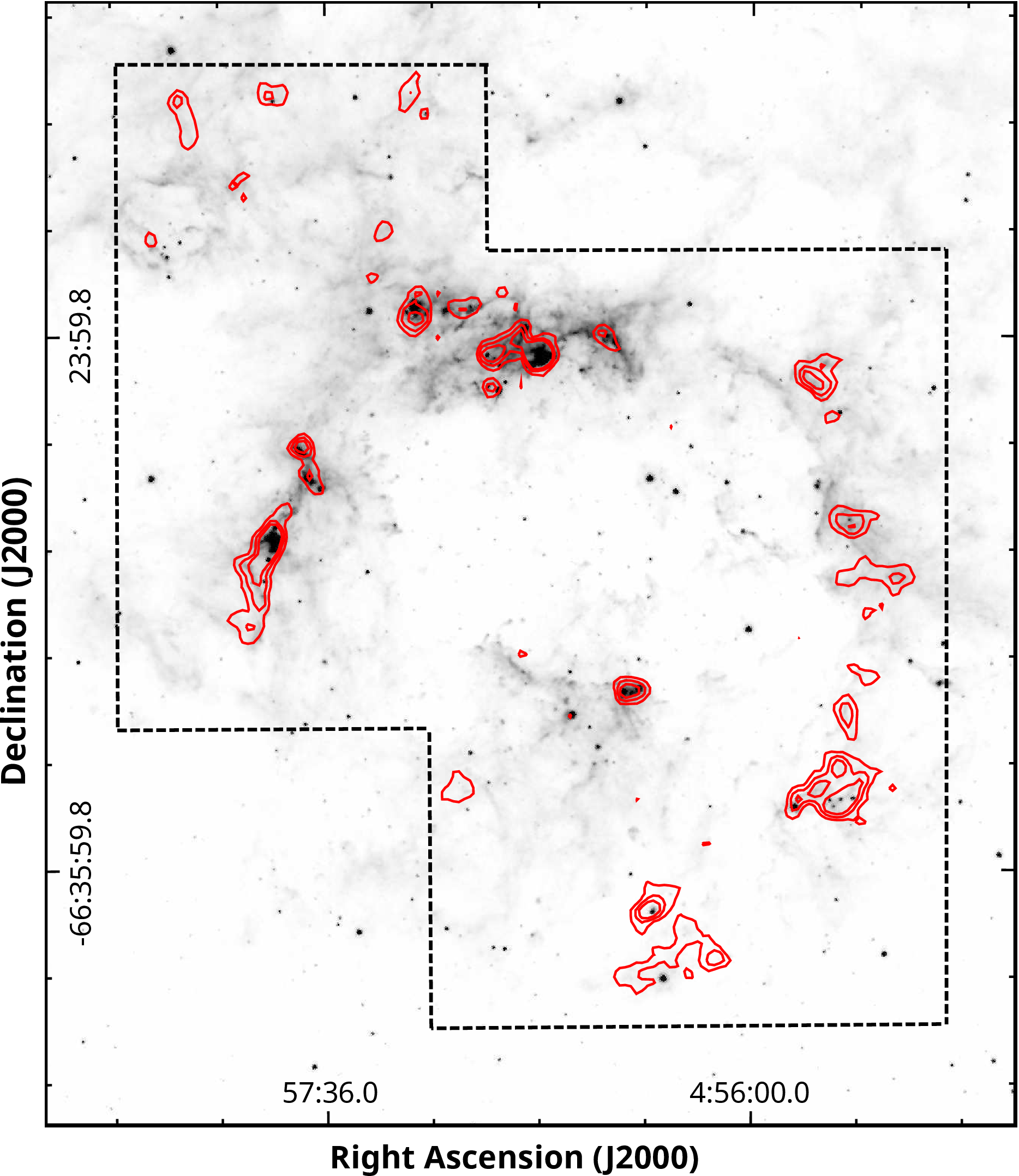}
  \caption{N11 complex as seen in the 8 $\mu$m emission obtained from SAGE Spitzer with the \2 J=3--2 emission integrated between 270 and 290 \ks~
presented in red contours. The contour levels are 4, 7, and 10  1.5 K \ks, and the rms noise level is between 1 and 1.5 K \ks.}
  \label{12comap}
\end{figure}

Figure\,\ref{12comap} presents in contours the \2 J=3--2 emission integrated between 270 and 290 \ks~toward 
the entire N11 region (the surveyed region is enclosed with the dashed lines).

\begin{figure}[h]
\centering
\includegraphics[width=9cm]{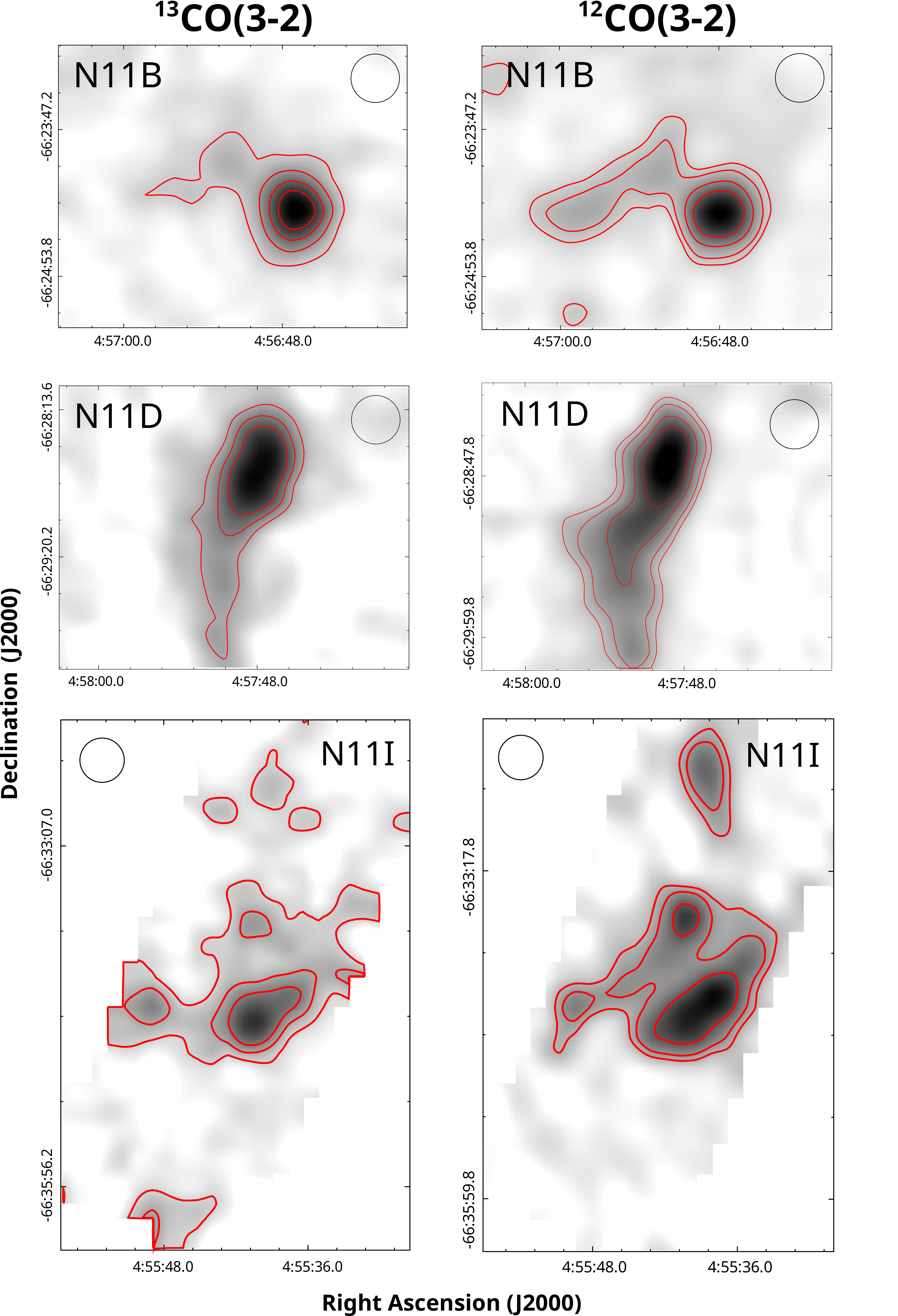}
\caption{Left: maps of the $^{13}$CO J=3--2 integrated emission between 270-290 \ks. The contour levels are 1, 2, 3 and 4 K \ks~for N11B (upper panel),
1, 1.5 and 2 K \ks~for N11D (middle panel) and 0.6, 1 and 1.5 K \ks~for N11I (bottom panel). The rms noise levels is about 0.2 K \ks. Right:
maps of the $^{12}$CO J=3--2 integrated emission between 270 and 290 \ks. The contour
levels are: 7, 10, 17 and 25 K \ks~for N11B (upper panel), 5, 8, 11 and 15 K \ks~for N11D (middle panel), and 5, 7 and 11 K \ks~for N11I
(bottom panel). The rms noise level is between 1 and 1.5 K \ks.
The beam is included in each panel and the rms noise levels is about 0.2 K \ks.}
\label{maps}
\end{figure}

Concerning the sub-regions mapped in the \3 (boxes in Fig.\,\ref{im_intro}), named N11B, N11D, and N11I, Fig.\,\ref{maps} (left) presents 
the integrated \3 J=3--2 line between 270 and 290 \ks. For comparison we also show in Fig.\,\ref{maps} (right) the same sub-regions as seen in 
the integrated $^{12}$CO J=3--2 line emission.

\begin{figure}[h]
  \centering
  \includegraphics[width=9.1cm]{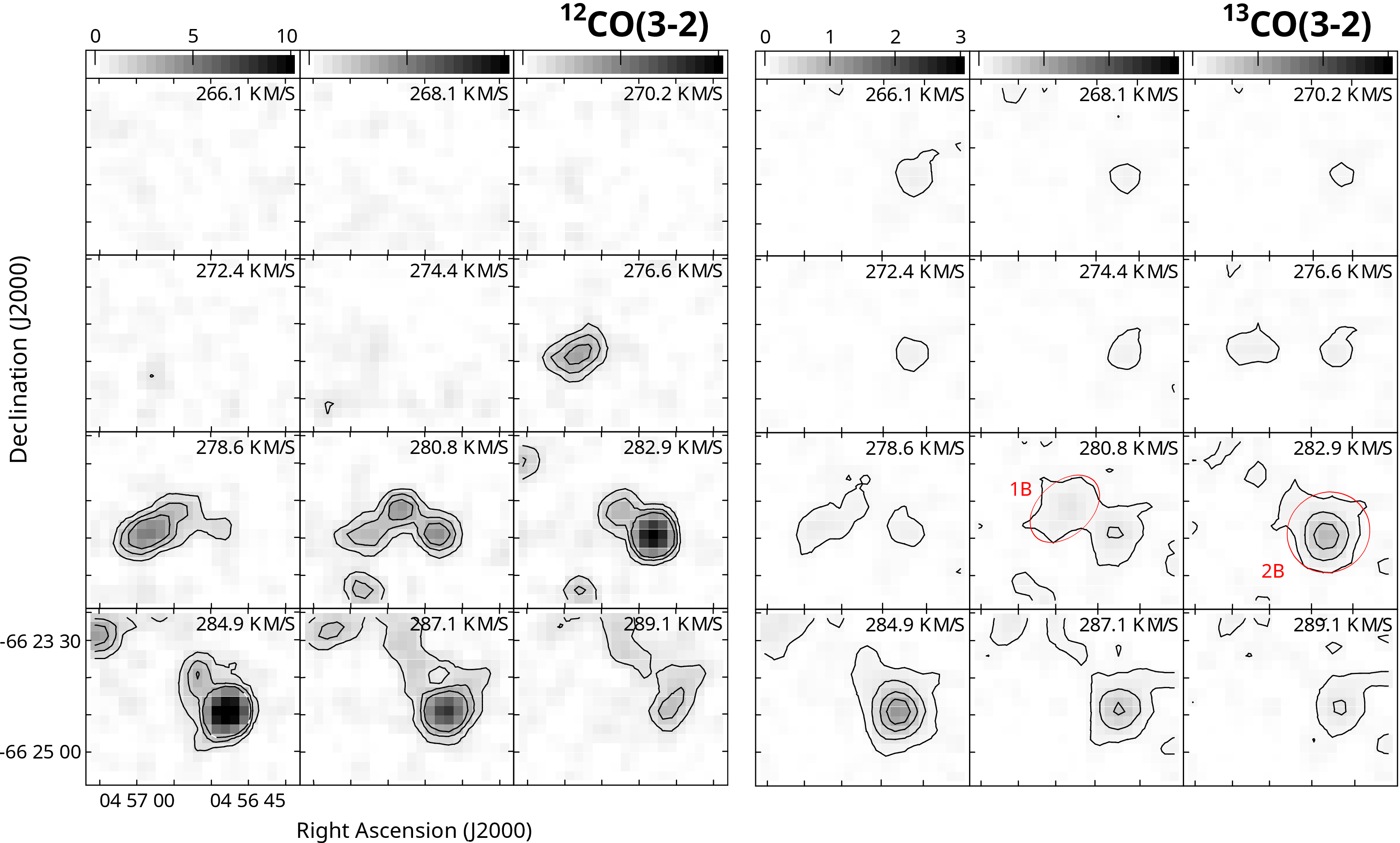}
  \caption{Channel maps of the $^{12}$CO and \3 J=3--2 line (left and right) of N11B.
The contour levels are 1, 2 and 3 K km s$^{-1}$ for the \2, and  0.1, 0.3, 0.6, and 0.9 K km s$^{-1}$ for the \3. 
The greyscale is displayed at the top of every map and is in K km s$^{-1}$. In two panels of the \3 emission, the positions
of two identified clumps are marked with red circles/ellipses. }
  \label{chanR1}
\end{figure}

\begin{figure}[h]
  \centering
  \includegraphics[width=9.1cm]{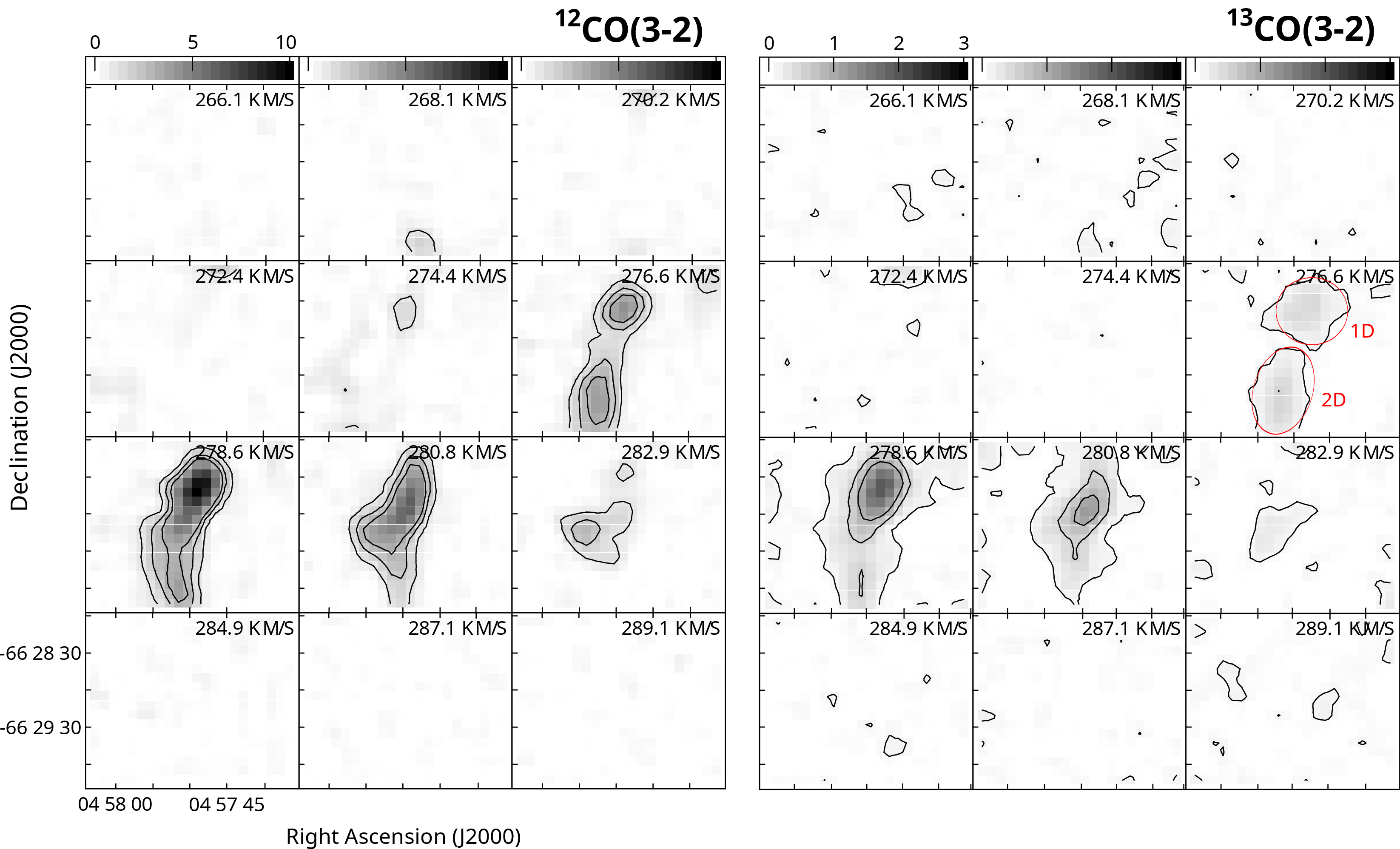}
  \caption{Channel maps of the $^{12}$CO and \3 J=3--2 line (left and right) of N11D.
The contour levels are 1, 2 and 3 K km s$^{-1}$ for the \2, and 0.1, 0.5 and 0.9 K km s$^{-1}$ for the \3. The greyscale is displayed at the 
top of every map and is in K km s$^{-1}$. In one panel of the \3 emission, the positions 
of two identified clumps are marked with red circles/ellipses. }
  \label{chanR2}
\end{figure}

\begin{figure}[h]
  \centering
  \includegraphics[width=9cm]{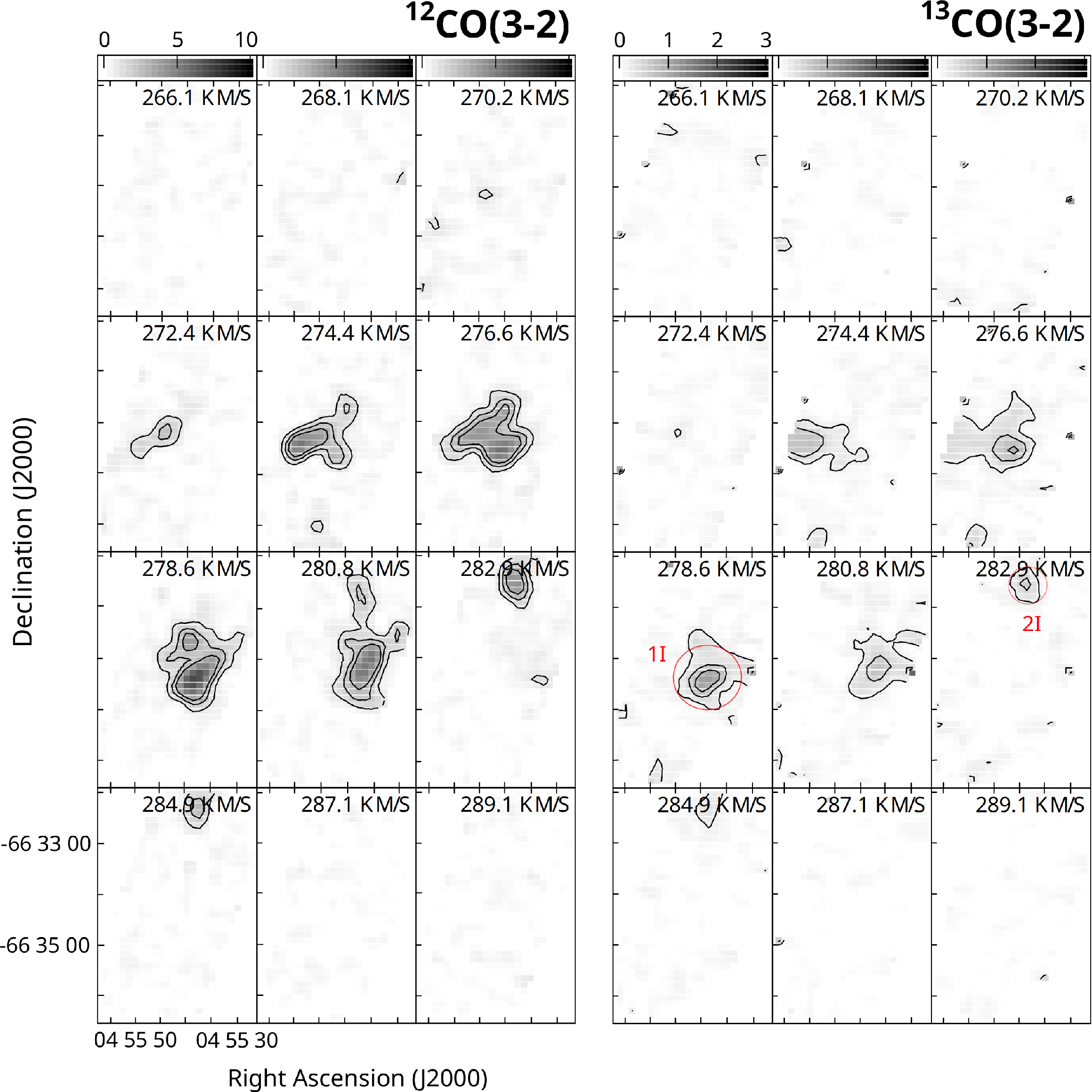}
  \caption{Channel maps of the $^{12}$CO and  \3 J=3--2 line (left and right) of N11I. 
The contour levels are 1, 2 and 3 K km s$^{-1}$ for the \2, and  0.2, 0.5 and 0.8 K km s$^{-1}$ for the \3. 
The greyscale is displayed at the top of every map and is in K km s$^{-1}$. In two panels of the \3 emission, the position 
of two identified clumps are marked with red circles. }
  \label{chanR3}
\end{figure}

In order to study the morphology and velocity distribution of the molecular gas mapped in each isotope toward N11B, N11D and N11I, 
we present in Figs.\,\ref{chanR1},\,\ref{chanR2},and\,\ref{chanR3} the \2 and \3 J=3--2 emission displayed in channel maps 
integrated in bins of 2 km s$^{-1}$. The velocity ranges 
displayed in these figures are the same as those presented in the \2~J=2--1 channel map in \citet{herrera2013}. We  
based our work on the \3 emission, the optically thinner tracer, to identify molecular clumps. Thus, in N11B we identified two clumps, 1B and 2B
(they are marked in the channel maps at 280.8 and 282.9 \ks~in Fig.\,\ref{chanR1}), in N11D we identified clumps 1D and 2D (marked in the channel 
maps at 276.6 \ks~in Fig.\,\ref{chanR2}), and finally in N11I, clumps 1I and 2I (marked in the channel maps at 278.6 and 282.9 \ks~in
Fig.\,\ref{chanR3}). In the last case, another molecular clump can be identified toward the east in the channel map at 274.4 \ks, but
as the clump is not completely observed we do not catalogued in order to calculate its physical parameters.

Taking into account that we have for the first time data of the \3 J=3--2 line toward N11, we use it to estimate the mass of 
the molecular clumps identified in Figs.\,\ref{chanR1}, \,\ref{chanR2}, and\,\ref{chanR3} by assuming local thermodynamic equilibrium (LTE). 
The excitation temperature was obtained from:

\begin{equation}
	T_{ex}(3\rightarrow2) = \frac{16.59 \textrm{K}}{\textrm{ln}[1+16.59 \textrm{K}/(T_{max}(^{12}\textrm{CO})+0.036 \textrm{K})]}
\end{equation}

\noindent The $^{12}$CO and $^{13}$CO optical depths, $\tau_{12}$ and $\tau_{13}$, were obtained from:

\begin{equation}
	\frac{^{12}\textrm{T}_{\textrm{mb}}}{^{13}\textrm{T}_{\textrm{mb}}}= \frac{1-\textrm{exp}(-\tau_{12})}{1-\textrm{exp}(-\tau_{12}/X)}
\end{equation}

\noindent where $^{12}$T$_{\textrm{mb}}$ and $^{13}$T$_{\textrm{mb}}$ are the peak temperatures of the $^{12}$CO and $^{13}$CO J=3--2 
lines and $X=50$ is the assumed isotope abundance ratio \citep{wang2009}. As shown in Table \ref{params}, 
the $^{13}$CO J=3--2 line is optically thin, thus we estimate its column density from:

\begin{equation}
	N(^{13}\textrm{CO})=8.28\times10^{13}e^{\frac{15.85}{T_{ex}}}\frac{T_{ex}+0.88}{1-e^{\frac{-15.85}{T_{ex}}}}\frac{1}{J(T_{ex})-J(T_{BG})}\displaystyle\int T_{mb}dv
\end{equation}

\noindent with 

\begin{equation}
	J(T)=\frac{h\nu/k}{exp\left(\frac{h\nu}{kT}\right)-1}
\end{equation}

We assume an abundance ratio of [H$_2$/$^{13}$CO] $= 1.8\times10^6$ \citep{garay2002} 
to obtain the molecular hydrogen column 
density N(H$_2$). Finally, the mass was derived from

\begin{equation}\label{masa}
	M=\mu m_H  D^2 \Omega  \sum_i{\rm N}_i({\rm H_2})
\end{equation}

\noindent where $\mu$ is the mean molecular weight, assumed to be 2.8 by taking into account a relative helium abundance of 25$\%$, 
$m_H$ is the hydrogen mass, $D$ is the distance (50 kpc) and $\Omega$ is the solid angle subtended by the beam size. 
We summed over all beam positions of the $^{13}$CO molecular structures. The central velocity, and the radii are presented in the first two 
lines in Table\,\ref{params}. In the case of clumps 1B and 2D, as they have an
elliptical shape, we provide the effective radius obtained from R$=(R_{a}\times R_{b})^{1/2}$, where $R_{a}$ and $R_{b}$ are
the semi-axes of the elliptical molecular feature.  The other files of the Table
present all the parameters obtained under the LTE assumption.

\begin{table}
\caption{Observed and derived parameters (assuming LTE) of each molecular clump.}
\label{params} 
\tiny
\centering
\begin{tabular}{l c c c c c c}
\hline\hline 
Clump	   	       &  1B    &  2B  & 1D    	   &   2D      & 1I      &  2I	 	\\
\hline  
Velocity [\ks]         &  278.5      &  284.0    &   278.5        &   279.5        &  278.5       &   283.5            \\
R [pc]                 &  6.4   &  6.5    &   7.7   & 5.3   & 8.4 & 6.3    \\
$\tau_{12}$	       &  8.8   &  9.5    & 10.6      & 	15       & 8.8     &  8.2        	 \\
$\tau_{13}$ 	       &  0.1   &  0.2    &  0.2      &   0.3        & 0.2     &     0.1  	 \\
T$_{ex}$ [K]           &  8	&  13    &  11	   &      7     &  10	 &       8        \\
N(H$_{2}$) [$\times10^{22}$~cm$^{-2}$] & 1.6        & 2.3       & 2.5    & 2.3	 & 3.9  & 1.0	\\
M$_{\rm LTE}$ [$\times10^4$~M$_{\odot}$]	       & 0.8         & 1.1       & 1.3    & 1.1      & 2.0    &	0.5  	\\
\hline
\end{tabular}
\end{table}

We estimate the virial mass of each clump from the \3 J=3--2 emission from the following equation:
\begin{equation} 
\rm  M_{vir}/M_{\odot}  = k~R/pc~(\Delta v_{13}/km~s^{-1})^{2}
\end{equation}
where k=190 by assuming clouds with density distributions $\propto$$r^{-1}$ \citep{mac88}, $\Delta$v$_{13}$ 
the line velocity width (FWHM) of the \3 J=3--2 emission obtained at the peak position of each molecular structure,
and R is the radius presented in Table\,\ref{params}. The $\Delta$v$_{13}$ and the obtained virial mass are presented 
in Table\,\ref{tvir}.

\begin{table}
\caption{Used $\Delta$v$_{13}$ (in \ks) and obtained virial mass (in $\times10^{4}$ M$_{\odot}$) of the molecular clumps from the \3 J=3--2 line.}
\label{tvir}
\tiny
\centering
\begin{tabular}{c c c c c c c}
\hline\hline
                                           & 1B   &  2B     &   1D   & 2D  & 1I  & 2I           \\
\hline
$\Delta$v$_{13}$                 &  4.0$\pm0.5$     &  5.4$\pm0.7$  &   3.7$\pm0.4$   & 4.5$\pm0.5$     & 6.0$\pm1.2$     & 2.6$\pm0.5$      \\
M$_{\rm vir}$                    &  1.9$\pm0.7$    &  3.7$\pm1.4$   &   1.9$\pm0.7$   &  2.0$\pm0.8$  & 5.7$\pm2.2$ &  0.8$\pm0.3$       \\
\hline
\end{tabular}
\end{table}

\begin{table}
\caption{Average integrated ratios toward the peaks of the molecular clumps. The errors in all cases are about 10\%.}
\label{ratios}
\centering
\begin{tabular}{l c c c c}
\hline\hline
 Clump & $\frac{^{12}\textrm{CO}(3-2)}{^{13}\textrm{CO}(3-2)}$   & $\frac{^{12}\textrm{CO}(3-2)}{^{12}\textrm{CO}(1-0)}$ & $\frac{^{12}\textrm{CO}(3-2)}{^{12}\textrm{CO}(2-1)}$\\
\hline
   1B   &   8.3   &    0.7         &   0.7          \\
   2B   &   7.2   &    1.2         &   0.9     \\ 
   1D   &   6.5   &    0.5         &   0.8     \\
   2D   &   6.8   &    0.7         &   0.7                     \\
   1I   &   7.0   &    0.4         &   0.6     \\
   2I   &   13.0  &    0.4         &   0.8                    \\
\hline
\end{tabular}
\end{table}

\begin{figure}[h]
  \centering
  \includegraphics[width=9cm]{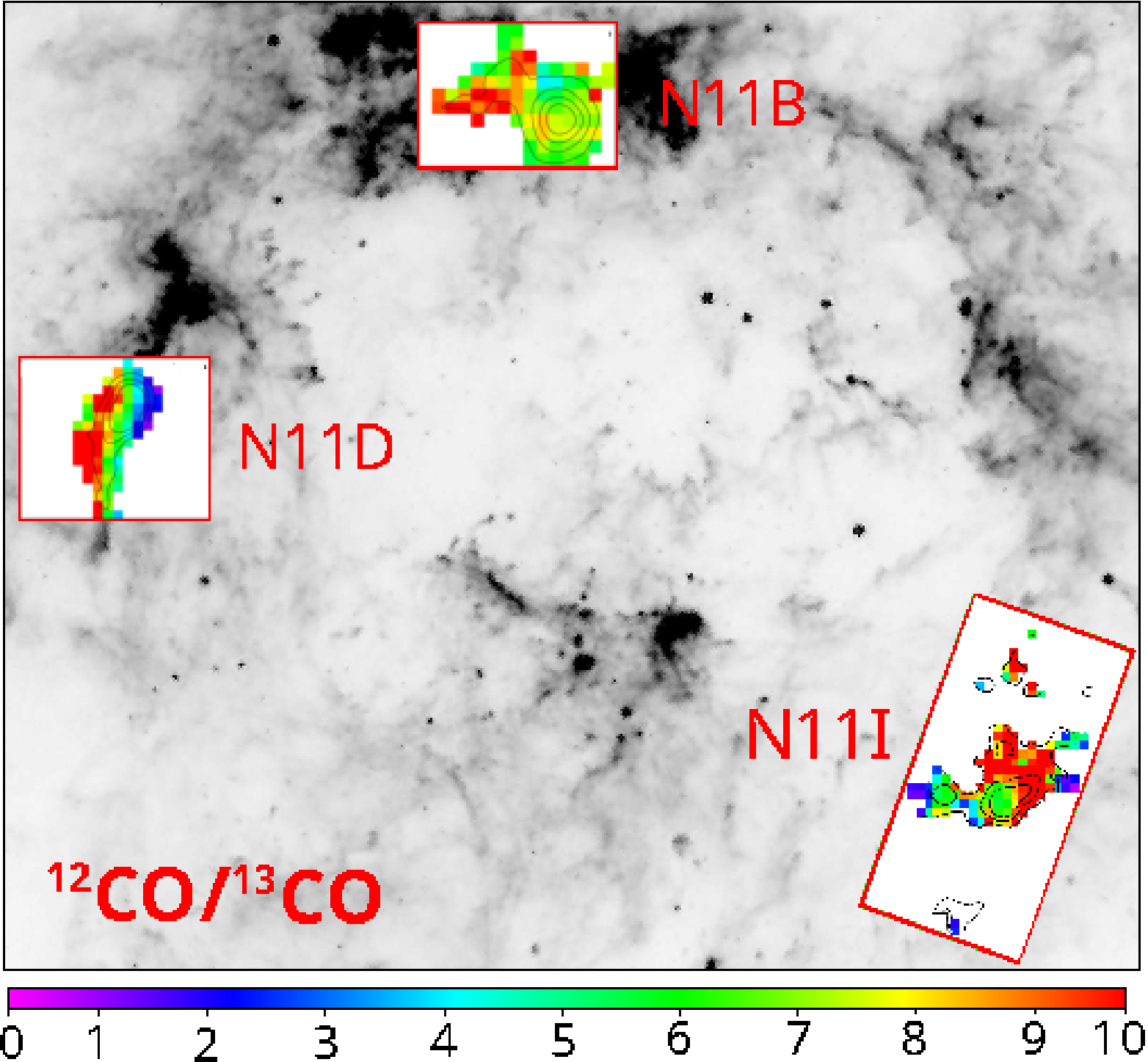}
  \caption{Isotopic integrated line ratio. The contours correspond to the integrated \3 J=3--2 emission as presented
in Fig.\,\ref{maps} (left).}
  \label{ratiomaps}
\end{figure}

\begin{figure}[h]
  \centering
  \includegraphics[width=9cm]{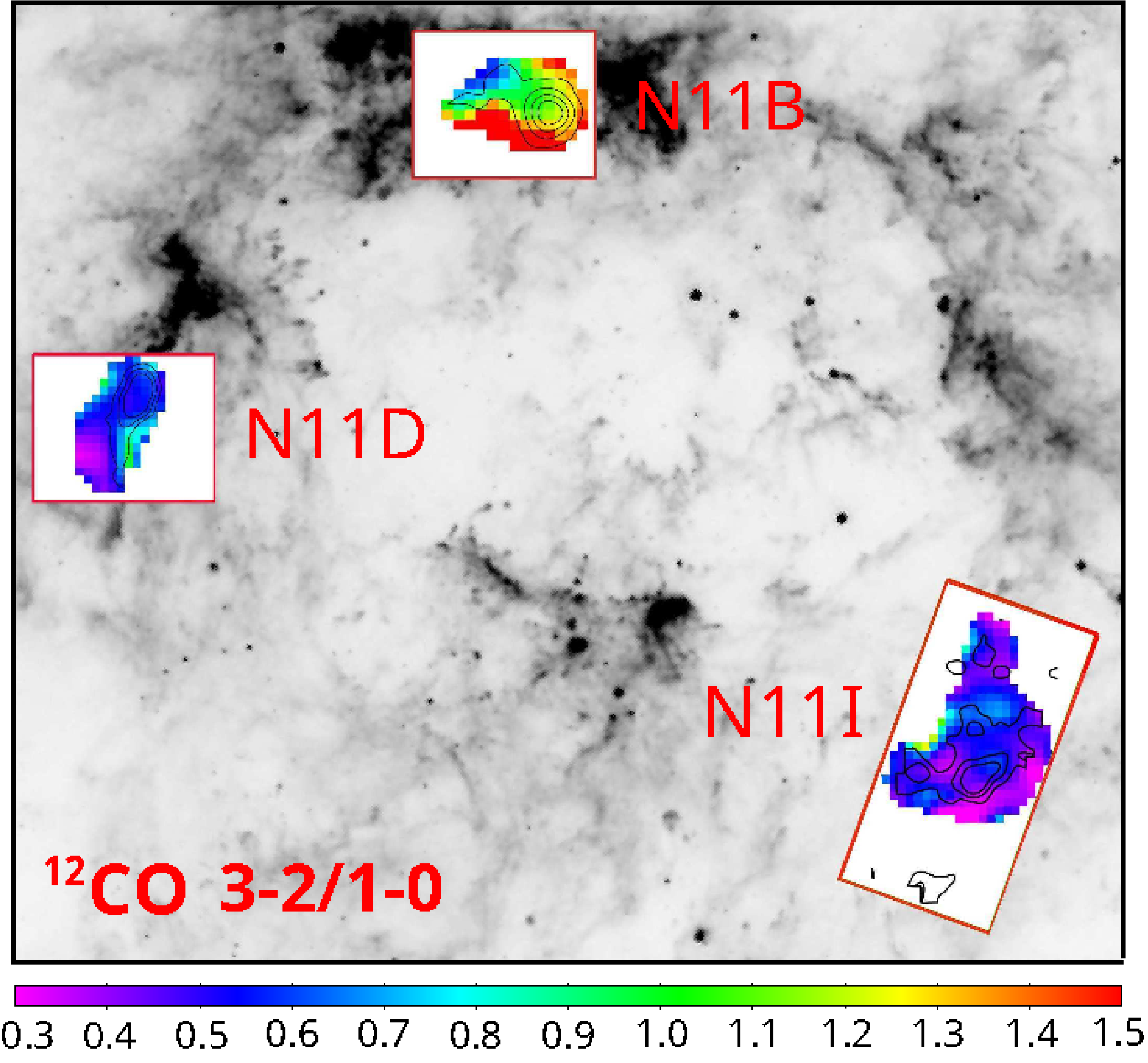}
  \caption{\2 J=3--2/1--0 integrated line ratio. The contours correspond to the integrated \3 J=3--2 emission as presented
in Fig.\,\ref{maps} (left).}
  \label{ratiomaps2}
\end{figure}

\begin{figure}[h]
  \centering
  \includegraphics[width=9cm]{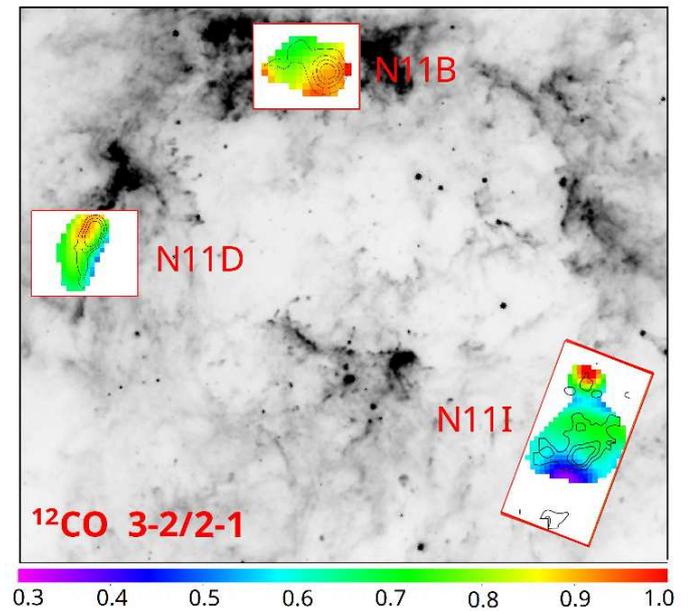}
  \caption{\2 J=3--2/2--1 integrated line ratio. The contours correspond to the integrated \3 J=3--2 emission as presented
in Fig.\,\ref{maps} (left).}
  \label{ratiomaps3}
\end{figure}

\subsection{Line ratio maps}

In order to study evidences of molecular photodissociation we calculate
the isotopic integrated line ratios $\int{\rm T^{^{12}CO}dv}/\int{\rm T^{^{13}CO}dv}$ (\2/\3 J=3--2).
A similar analysis was done for the $\int{\rm T^{3-2}dv}/\int{\rm T^{2-1}dv}$  (\2 3--2/2--1) and
$\int{\rm T^{3-2}dv}/\int{\rm T^{1-0}dv}$ (\2 3--2/1--0) ratios in order to study possible evidences of shocks in the gas, 
how the gas is affected by the radiation, and to better constrain the relation between the J=3--2 and J=1--0 lines in order to
evaluate CO contribution to the continuum at 870 $\mu$m.

We construct maps of \2/\3 J=3--2 
for each region, which are presented in Fig.\,\ref{ratiomaps} in the context of the N11 complex. To construct the maps of the 
\2/\3 ratio it was considered pixels above the 3$\sigma$ noise level in both, the \3 and \2 integrated maps.  

Using the \2 J=2--1 and J=1--0 from \citet{herrera2013} together with our observations of \2 J=3--2, we construct 
maps of the integrated line ratios \2 3--2/2--1 and \2 3--2/1--0  (see Figs.\,\ref{ratiomaps2} and\,\ref{ratiomaps3}). 
To compare these lines, we need to have the data at the same angular resolution. 
Thus, we convolve the $^{12}$CO J=3--2 data to the 23\arcsec~and 45\arcsec~resolutions of 
the J=2--1 and J=1--0 data, respectively. 

Table\,\ref{ratios} lists the obtained average values for these ratios toward the peak position of 
each molecular clump.

\subsection{Non-LTE analysis}

Using the \2 J=1--0, 2--1, and 3--2 and \3 J=3--2 lines we perform a non-LTE analysis to estimate the physical conditions toward
the peaks of the most intense molecular features (i.e. clouds 2B, 1D, and 1I). 
To estimate the column density and the H$_{2}$ volume density, we use the RADEX code\footnote{RADEX is a statistical equilibrium 
radiative transfer code, available as part of the Leiden Atomic and Molecular Database (http://www.strw.leidenuniv.nl/moldata/).} \citep{tak2007}.

\begin{figure*}[tt]
  \centering
  \includegraphics[width=17cm]{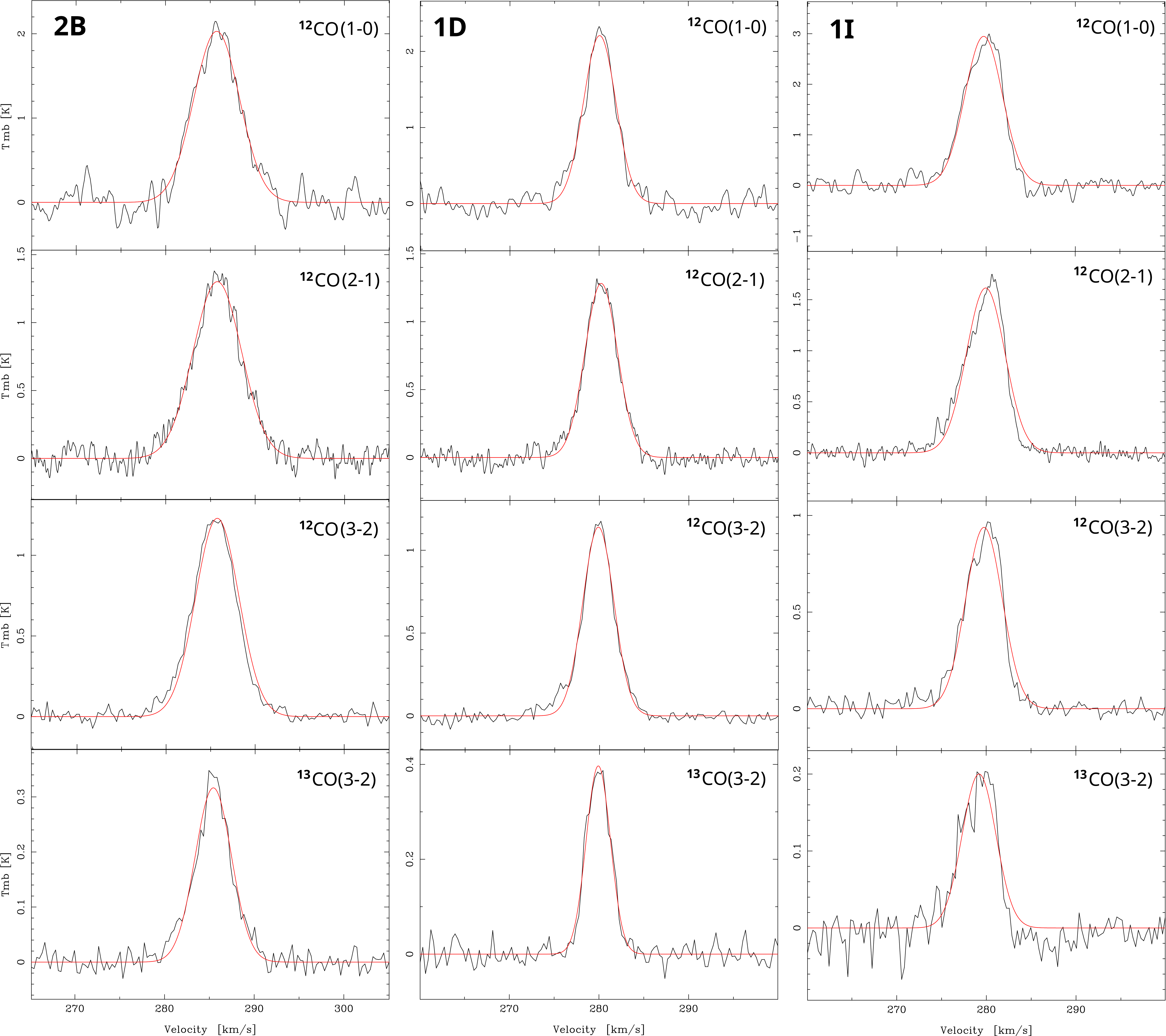}
  \caption{\2 J=1--0. 2--1, and 3--2, and \3 J=3--2 spectra toward the peak of clump 2B, 1D, and 2I. The red curves are the Gaussian fits.
The rms noise of each spectrum is (from top to bottom): 178.0, 95.2 and 31.2 mK (clump 2B), 140.0, 62.9 and 25.6 mK  (clump 1D), 
and 142.0, 55.7 and 35.6 mK (clump 2I). The \2 J=2--1 and 3--2, and \3 J=3--2 spectra were convolved with the 45\arcsec~resolution of
the J=1--0 line.}
  \label{specR}
\end{figure*}

The inputs of RADEX are: kinetic temperature (T$_{\rm K}$), line velocity width at FWHM ($\Delta$v), and
line peak temperature (T$_{\rm peak}$), and the code yields column and volume molecular densities (N and n$_{\rm H_{2}}$).
Assuming that the dust and gas are coupled, we use T$_{\rm K} = 20$ K based on the dust temperature obtained in previous 
works \citep{galametz16,herrera2013}.
In the case of the \2 J=2--1 and 3--2 and \3 J=3--2 lines, the data were convolved with the 45\arcsec~beam of the J=1--0 line. 
The T$_{\rm peak}$ and $\Delta$v were obtained from Gaussian fits to the spectra (see Fig.\,\ref{specR}),
whose results are presented in Table\,\ref{fitsTable}. Results from the \3 J=3--2 line are included for comparison.

\begin{figure}[h]
  \centering
  \includegraphics[width=9cm]{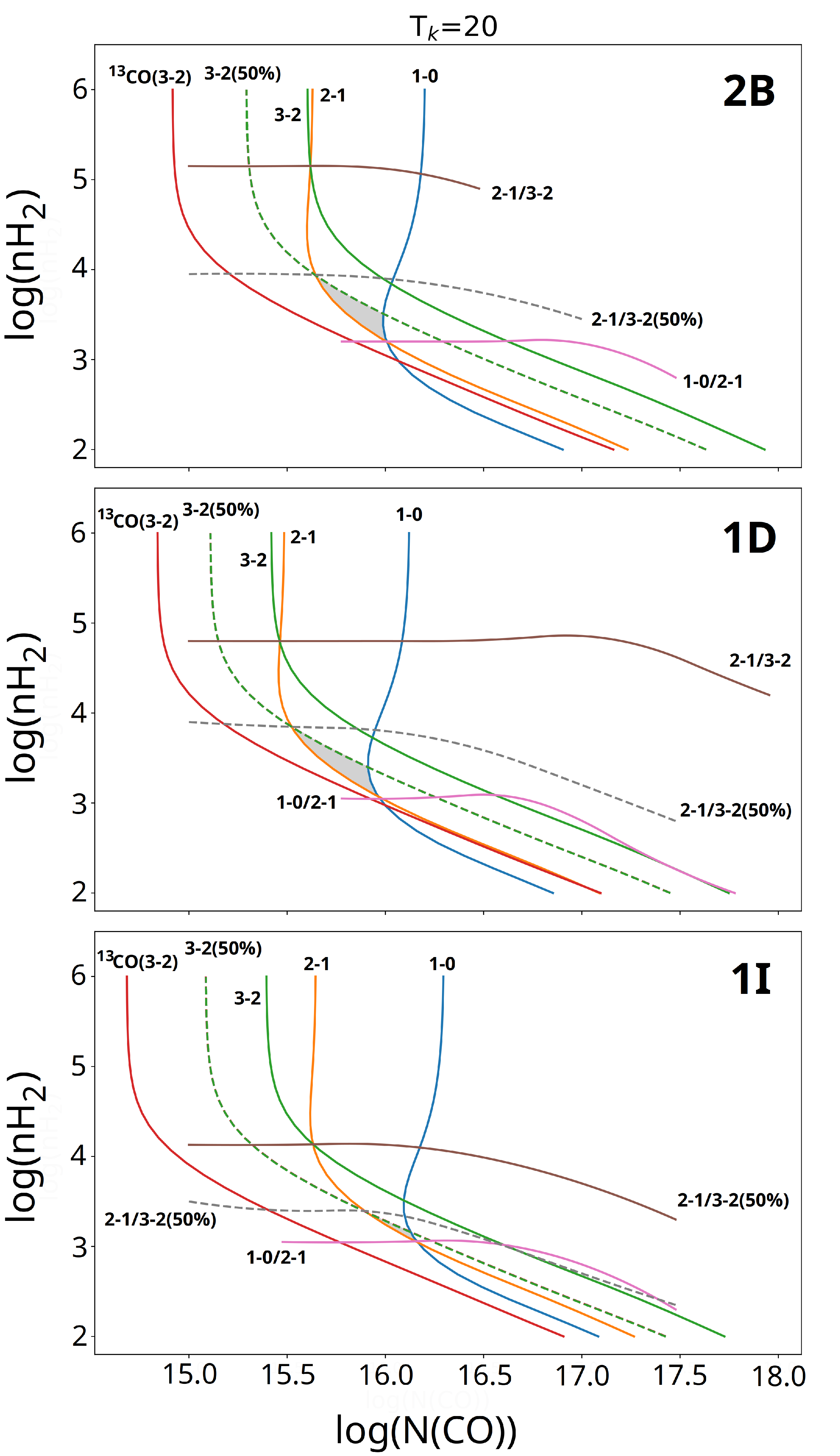}
  \caption{Radex results obtained from the \2 lines toward the peaks of the most intense clumps: 2B, 1D, and 1I (from top to bottom).
Dashed lines are the results obtained from the assumption that the 50\%~of the \2 J=3--2 emission arises from the cold component at 20 K.
The gray shaded areas in each panel show the regions of most probable values of N(CO) and n$_{\rm H_{2}}$.
For comparison, results obtained from the \3 J=3--2 line is also included. }
  \label{radexFig}
\end{figure}

As done in a previous work toward N159 \citep{paron2015}, we assume that the lower
\2 transitions (J=1--0 and 2--1) arise mainly from the cold gas component, while it is likely that the
J=3--2 transition arises from both, a cold and a warmer one. Then, we run the RADEX code for two cases: 100 and 50 per cent of 
the \2 J=3--2 emission assigned to the cold component at 20 K. The results from RADEX are presented in 
Figure\,\ref{radexFig}. Table\,\ref{tab_BDI} presents the results obtained by considering the 50 per cent of  
the \2 J=3--2 emission, which as Fig.\,\ref{radexFig} shows (see the gray shaded area in each panel) are the more tight results 
between the analyzed possibilities.
Taking into account the errors in the peak temperatures, the uncertainties in the results from RADEX are about 20\%~and 30\%~in 
the limits of the ranges of N(CO) and n$_{\rm H_{2}}$, respectively.

\begin{table}
\caption{Line parameters from the spectra shown in Figs.\,\ref{specR}. The \2 J=2--1 and 3--2, and \3 J=3--2 spectra were convolved 
with the 45\arcsec~resolution of the J=1--0 line. }
\label{fitsTable}
\centering
\begin{tabular}{ccccc}
\hline\hline
Clump  & Emission   &  $T_{\textrm{mb}}$  &   $V_{\textrm{LSR}}$  &   $\Delta$ v      \\
        &           &  (K)                &    (km s$^{-1}$)       &   (km s$^{-1}$)   \\
\hline
2B      &  \2 (1--0) &     2.0 $\pm$ 0.2    &      285.8 $\pm$ 0.3  &      5.9 $\pm$ 0.1   \\
        &  \2 (2--1) &     1.3 $\pm$ 0.1    &      285.8 $\pm$ 0.1  &      6.2 $\pm$ 0.7   \\
        &  \2 (3--2) &     1.2 $\pm$ 0.3    &      285.4 $\pm$ 0.7  &      5.7 $\pm$ 1.7   \\
        &  \3 (3--2) &     0.30 $\pm$ 0.03  &      285.4 $\pm$ 1.0  &      4.8 $\pm$ 0.6    \\
\hline
1D      &  \2 (1--0) &     2.3 $\pm$ 0.2    &      280.1 $\pm$ 0.2  &      4.5 $\pm$ 0.5   \\
        &  \2 (2--1) &     1.3 $\pm$ 0.1    &      280.2 $\pm$ 0.3  &      4.3 $\pm$ 0.6   \\
        &  \2 (3--2) &     1.1 $\pm$ 0.4    &      279.9 $\pm$ 0.7  &      4.1 $\pm$ 0.8   \\
        &  \3 (3--2) &     0.40 $\pm$ 0.03  &      279.9 $\pm$ 1.0  &      3.1 $\pm$ 0.4    \\
\hline
1I      &  \2 (1--0) &     2.9 $\pm$ 0.9    &      279.7 $\pm$ 0.2  &      4.9 $\pm$ 0.4   \\
        &  \2 (2--1) &     1.6 $\pm$ 0.1    &      279.9 $\pm$ 0.2  &      5.1 $\pm$ 0.5   \\
        &  \2 (3--2) &     0.9 $\pm$ 0.3    &      279.8 $\pm$ 0.7  &      4.8 $\pm$ 1.5   \\
        &  \3 (3--2) &     0.20 $\pm$ 0.03  &      279.1 $\pm$ 0.6  &      4.4 $\pm$ 0.9   \\
\hline
\end{tabular}
\end{table}

\begin{table}
\caption{Radex results from the $^{12}$CO 1--0 , 2--1 and 3--2(50\%) lines with T$_{\rm K} = 20$K. Errors are about 20\%~and 30\%~in
the limits of the ranges of N(CO) and n$_{\rm H_{2}}$, respectively. }
\label{tab_BDI} 
\centering
\begin{tabular}{l l l}
\hline\hline 
Clump  &  	N(CO) ($10^{15}$cm$^{-2}$) &	n$_{\rm H_{2}}$ ($10^3$cm$^{-3}$)   \\
\hline   
2B    	 &  4.4 -- 10.1      & 1.5 -- 8.5				\\
1D       &  3.3 -- 9.1       & 1.1 -- 6.9                                \\      
1I       &  7.9 -- 14.4      & 1.0 -- 2.4				\\
\hline
\end{tabular}
\end{table}

\section{Discussion}

The \2 and \3 J=3--2 line were mapped for the first time toward N11. These data complement previous 
studies about the molecular gas and dust properties in the region (e.g. \citealt{israel03,herrera2013,galametz16}). 
The \2 J=3--2 emission shows that the molecular gas is distributed in a fragmented shell around N11 (see Fig.\,\ref{12comap}).
As \citet{deha05} point out, the presence of either a dense molecular shell surrounding the
ionized gas of an \hii~region or massive fragments regularly spaced along the ionization front, suggests triggered 
star formation through the collect and collapse mechanism. This scenario agrees 
with sequential star formation among the OB associations proposed in \citet{parker1992} and \citet{walb92}, and suggests that it is 
likely that the triggered star formation processes observed in our Galaxy (e.g. \citealt{deha09,zav10,duro17}) also occur in the LMC.

The morphology of the molecular features, as seen in the \2 and \3 J=3--2 lines 
(see Fig.\,\ref{maps} and also the channel maps in Figs.\,\ref{chanR1},\,\ref{chanR2},and\,\ref{chanR3}), 
shows signatures that allow us to discuss some issues about the effects of 
the \hii~regions/OB associations on the molecular gas. For instance, in the case of N11D, which is a region southwards the LH13 OB association, 
the observed molecular cloud has a filament-like structure with a maximum toward the north. This feature presents a slight curvature
with the concavity pointing to the center of N11 nebula. Even though this cloud can be affected by the action of LH13 OB association, 
taking into account its morphology and that it is located at a border of the N11 complex, the ionizing bubble around LH9 
(the N11 central OB cluster) should be important in shaping it.

In general, the analyzed molecular features in the regions that have been observed in \3 show a clumpy morphology with at least two well
defined molecular clumps in each one. In the case of N11D we identified clumps 1D and 2D belonging to the molecular 
structure described above.
In N11B, there are two molecular structures: a more intense one with a circular
shape (clump 2B), and a weaker one with a roughly elliptical shape (clump 1B). These molecular
features positionally coincide with the nebula generated by LH10 and clump 2B is located slightly to the north of a maximum of [OIII] emission
generated by the LH10 OB association, where also a methanol maser lies \citep{barba03}. 
The molecular feature in N11I presents an even more clumpy
morphology, with clump 1I being the more intense one in the region. The presence of these clumps in the \3 J=3--2 line can be interpreted as an 
evidence of the 
fragmentation generated in a molecular shell due to the expansion of the N11 nebula as mentioned above. However, a general collapse of 
giant molecular clouds giving place to these molecular fragments can not be discarded.

From the mass characterization of these clumps we find that the ratio between the virial and LTE mass (M$_{\rm vir}$/M$_{\rm LTE}$)
is higher than unity in all cases, which suggests
that they are not gravitationally bounded, and it is likely that we are mapping mainly turbulent molecular gas. 
This is consistent with the presence of OB clusters and \hii~regions, known sources of kinetic energy and turbulence in molecular 
clouds (e.g. \citealt{garay2002}). \citet{herrera2013}, using the \2 J=2--1, obtained similar relations between 
M$_{\rm vir}$ and M$_{\rm LTE}$ for most of 
their catalogued clumps in N11. Our results, obtained with the \3, an optically thinner tracer, show that even deeper parts of the molecular clumps
are not gravitationally bounded and may be supported by
external pressure, as found in molecular clouds around Galactic \hii~regions \citep{rath02,massi97}, and in 
the LMC, in N113 \citep{paron2014} and Cloud B of Complex No.37 \citep{garay2002}. 

Clump 1I exhibits the largest virial mass among the studied clumps ($\sim 5.7 \times 10^4$ \msol) which 
is almost three times larger than its LTE mass. This suggests, as mentioned above, that the molecular clump would not be gravitationally supported. 
Taking a look to the corresponding $^{12}$CO J=1-0, J=2-1, and J=3-2 spectra (see Fig. \ref{specR}), a  
slight red-skewed asymmetry in the profiles is observed. Discarding the presence of a second velocity component, and given that it is very likely that
this clump lacks stellar activity in its interior, the observed red-skewed asymmetry in the spectra could be ascribed to the global 
expansion of N11 originated by the feedback of the OB association LH9. Taking into account that the V$_{\rm LSR}$ associated with the \2
emission is about 
280~kms$^{-1}$ and the average systemic velocity of the association LH9 is about 295 \ks~\citep{evans07}, the blue wings exhibited 
in the spectra could be reflecting the presence of gas pushed toward us by the expansion of the nebula. 

\subsection{Integrated line ratios}

The average values (between 7 and 8) in the \2/\3 integrated line ratio found toward the peaks of clumps 2B, 1D, 2D, and 1I 
are similar to the values found toward N159 and N113 \citep{paron2014,paron2015}, while the values at the peaks of clumps 2I and 1B 
(larger than 8) are quite similar to the values found toward N132, N166-A, and N166-B \citep{paron2015}, regions not as active as N133 and N159.
In general, the \2/\3 maps (see Fig.\,\ref{ratiomaps}) show that the molecular peaks present a quite uniform ratio, with 
larger values at the borders. 
By assuming that the integrated line ratios resemble the \2/\3 abundance ratio, the selective photodissociation of the \3 at the external 
layers of the clumps can explain
what is observed \citep{van88,visser09}. This phenomenon has been observed in Galactic molecular clouds exposed to 
UV radiation \citep{langer93,visser09}. 
In the case of N11D, a clear gradient is found extending from right to left, with lower values at the border exposed
to the N11 central OB cluster (LH9) and
larger values at the opposite one. This cannot be explained just by selective photodissociation
generated by a FUV flux from right to left, suggesting either the influence of an external source of flux impinging from left to right
in N11D such as a strong interstellar radiation field, or most likely, competing mechanisms
among selective photodissociation, chemical fractionation and non-isotope-selective reactions \citep{sz14,feder03}.

The average value of the \2 J=3--2/1--0 ratio obtained from the six analyzed molecular clumps is about 0.65, which is in agreement with the 
assumed value in \citet{galametz16} to estimate the contribution of the $^{12}$CO J=3--2 line to the 870 $\mu$m flux observed with LABOCA. The
bandwidth of LABOCA camera is wide and includes the \2 line, thus, it is contaminated and the contribution of the \2 emission to the
continuum flux has to be removed. Given that observations of \2 J=3--2 were lacked,
\citet{galametz16} used the observed $^{12}$CO J=1--0 line to estimate the emission of the J=3--2 line. Our results confirm
the estimation performed by the authors. However, considering the variations in the \2 J=3--2/1--0
ratio among the regions (see Fig.\,\ref{ratiomaps2}), the use of the \2 J=3--2 emission presented here could improve the estimation of the 
actual contribution of the dust emission at 870 $\mu$m. In Sect.\,\ref{dust} we discuss this contribution from a comparison
with the LABOCA data.

Additionally, from Fig.\,\ref{ratiomaps2}, we note that N11B presents \2 J=3--2/1--0 ratios higher than unity across most of the mapped area, 
which is in agreement with \citet{minami08}, who point out that this ratio is enhanced to 1.0 -- 1.5 
toward \hii~regions or clouds with young clusters, as it is the case for N11B. On the other hand, the authors found ratios less than 1 
toward clumps with neither \hii~regions nor clusters, as it is the case for N11I. N11D also presents
\2 J=3--2/1--0 ratios less than 1, suggesting that N11B and N11D may be affected in different ways by the radiation. 

The values of \2 J=3--2/2--1 ratio are quite constant among the peaks of the analyzed molecular clumps, with an average of 0.75. This value
is similar to what is found toward regions along molecular loops at the Central Molecular Zone in our Galaxy \citep{kudo11}; in particular
toward some protrusions in these features, that have larger values than in the surrounding gas, but lower than in strong shocked 
regions (with values as high as 2.5). Indeed, our values in the \2 J=3--2/2--1 ratios are close to the value (0.8) found in typical 
galactic disk clouds \citep{enokiya18}. The maps presented in Fig.\,\ref{ratiomaps3} show that the values in the \2 J=3--2/2--1 ratio
increase toward the south of N11B and the north of N11D, in coincidence with the presence of the local OB associations in these regions.

\subsection{\2 J=3--2 line contribution to the 870 $\mu$m continuum}
\label{dust}

\begin{figure}[h]
  \centering
  \includegraphics[width=6.5cm]{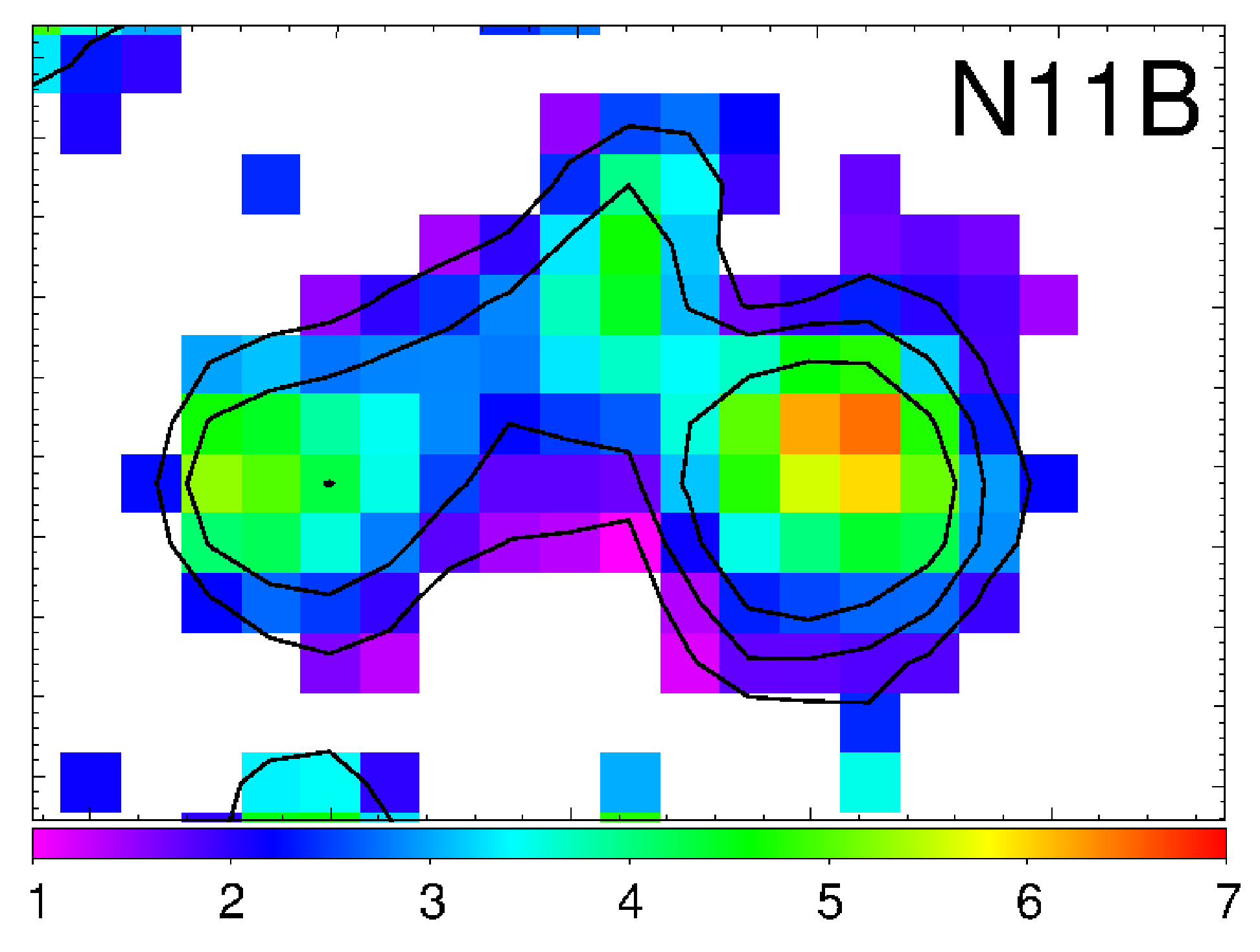}
  \includegraphics[width=6.5cm]{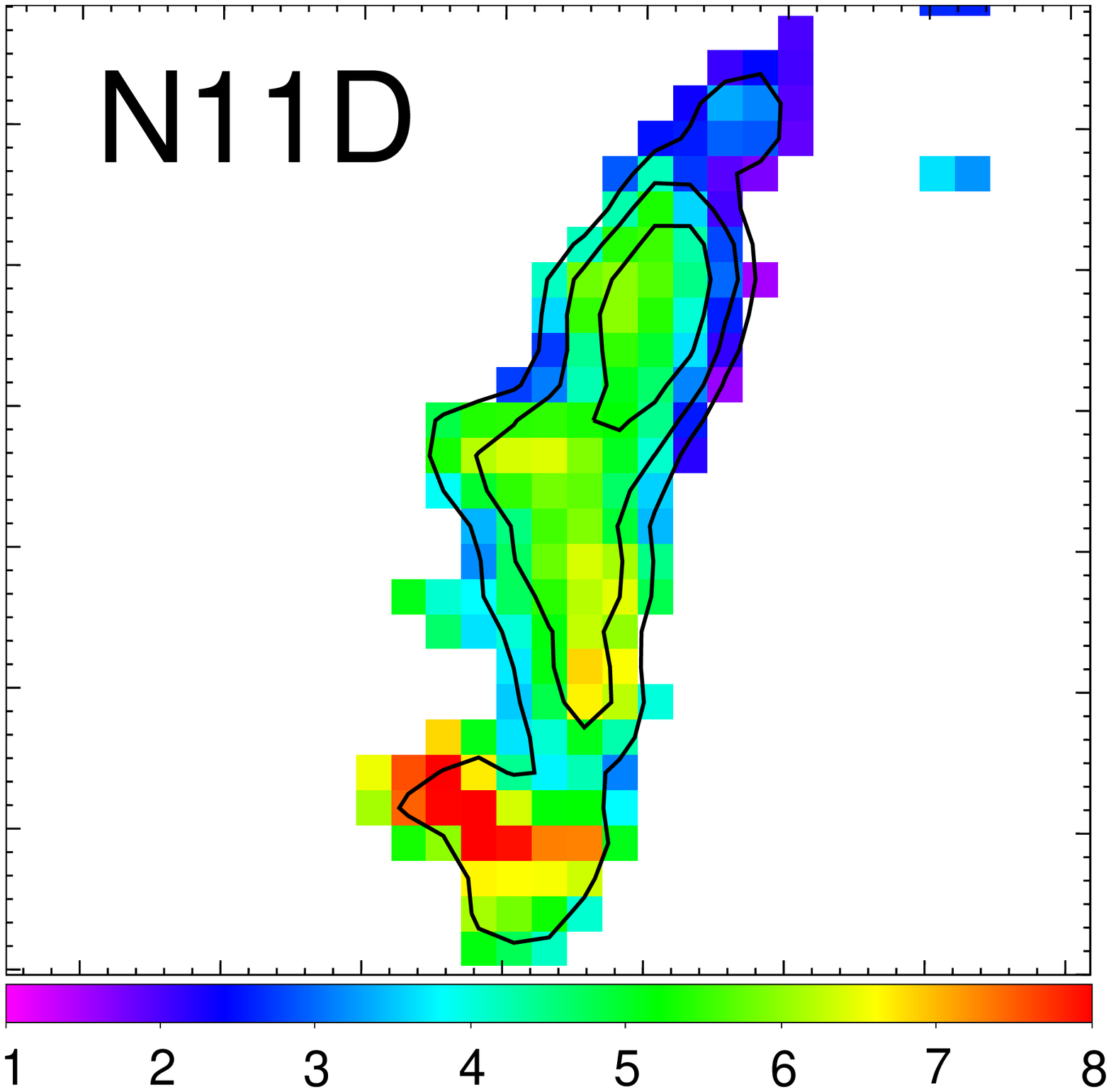}
  \includegraphics[width=6.5cm]{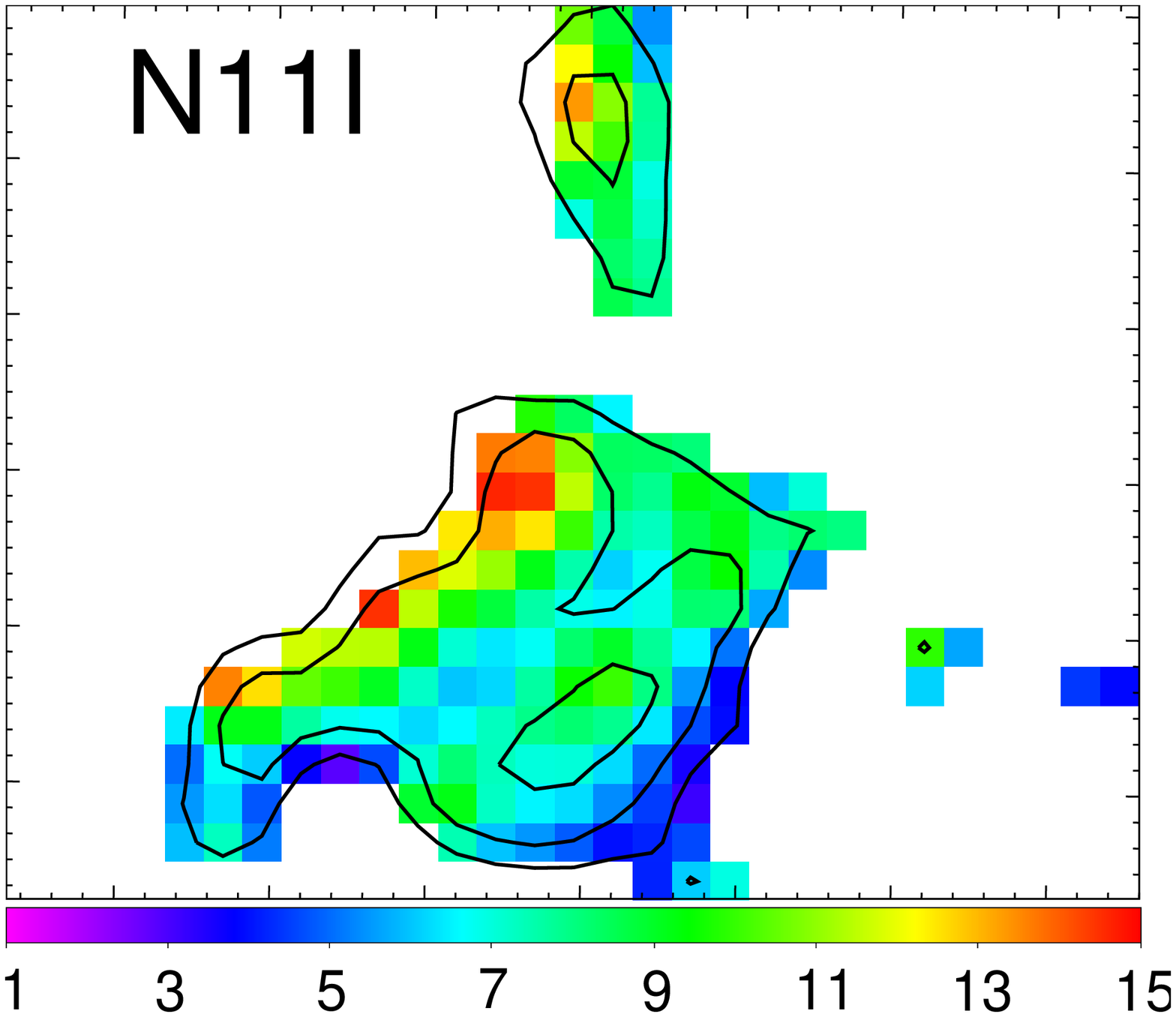}
  \caption{Contribution of the \2 J=3--2 emission to the the 870 $\mu$m continuum. The colobar represents the percentage (\%). Some contours
of the integrated \2 J=3--2 are included for reference.}
  \label{linepercent}
\end{figure}

Following the above discussion, we calculate the \2 J=3--2 line contribution to the continuum emission at 870 $\mu$m by 
comparing with LABOCA data (kindly provided by Galametz M.). The continuum data were convolved to the angular resolution
of the \2 data. The integrated \2 J=3--2 line (in K \ks) was converted to pseudo-continuum flux (mJy beam$^{-1}$) using
the conversion factor $C$ (see \citealt{drabek12}). 
The maps presented in Fig.\,\ref{linepercent} show the percentage of the line contribution to the 870 $\mu$m continuum toward the analyzed 
regions. The map of N11B shows similar values as presented in \citet{galametz16}. On the other hand, in the case of 
N11D, our map shows lower percentages in comparison with the range presented by the mentioned authors.

It is worth noting that region N11I presents the larger line contribution to the 870 $\mu$m continuum among the three studied
regions. This could be due to the fact that this region is less irradiated by UV photons and hence the molecular gas, in comparison with 
the dust, should be more abundant than in regions more intensely irradiated.

\subsection{Non-LTE considerations}

Concerning to the non-LTE analysis, a T$_{\rm K} = 20$ K was considered based on the assumption 
that the gas is coupled to the dust, and hence they should have same temperatures. We find that the best convergence in the 
models (i.e. obtaining ranges of results more tight in a N(CO)--n$_{\rm H_2}$ diagram) occurs when
we consider that the \2 J=3--2 emission likely arises from both, a cold and a warmer gas component. 
Thus, we roughly consider that 50\%~of its emission corresponds to gas with the assumed T$_{\rm K} = 20$ K while the other
50\%~to gas at higher temperatures.

RADEX yields n$_{\rm H_2}$ ranges of 1.5 -- 8.5, 1.1 -- 6.9, 1.0 -- 2.4  $\times$ 10$^{3}$ cm$^{-3}$ for clumps 2B, 1D, 
and 1I, respectively. The range with largest values in n$_{\rm H_2}$ is found toward 2B, 
which is the unique region with direct evidence of ongoing star formation, while a more narrow range with lower values is found toward 1I,
the region without any activity within it. The average value of n$_{\rm H_2}$ obtained in clump 2B, and the largest values
in clump 1D are in quite agreement with the density obtained toward the star-forming region N113 by \citet{wang2009} from a 
LVG analysis.

\section{Summary}

The N11 nebula, one of the most important star-forming region in the Large Magellanic Cloud, 
was mapped for the first time in the \2 and \3 J=3--2 line using the ASTE Telescope. 
The \2 J=3--2 line, mapped in the whole region, shows that the molecular gas is distributed in a fragmented shell around N11.
Three sub-regions (N11B, N11D, and N11I)  that  may be affected by the radiation in different ways, were also mapped in the \3 J=3--2 line. 
N11B and N11D are related to the OB associations LH10 and LH13, respectively, and 
N11I is a farther area at the southwest without any embedded OB association. The main results are summarized as follows:

(1) We found that the molecular features lying in each analyzed sub-region are in general clumpy (at least two well defined
clumps were found in each one) and show some signatures that could be explained by the expansion
of the nebula and the action of the radiation. In N11D, which is a region southwards the LH13 OB association, the molecular cloud has 
a curved filament-like morphology with the concavity pointing to the center of N11 nebula, showing that the ionizing bubble around 
LH9 (the N11 central OB cluster) seems to shape the N11D molecular cloud, influencing its star-formation and chemistry as suggested
by the isotopic ratio. The molecular cloud in N11B has
a peak slightly to the north of a maximum of [OIII] emission generated by the LH10 OB association.
The molecular feature in N11I, which is even more clumpy and it is not related to any embedded OB association, 
reinforces the hypothesis of the fragmentation generated in a large molecular shell due to the expansion of the N11 nebula. 

(2) The fragmented molecular clouds observed along the N11 bubble are in line with the sequential star formation scenario 
proposed by previous works but also observed in our Galaxy. 

(3) The ratio between the virial and LTE mass (M$_{\rm vir}$/M$_{\rm LTE}$) is higher than unity in all analyzed molecular clumps, 
which suggests that they are not gravitationally bounded and may be supported by external pressure.

(4) A non-LTE analysis suggests that at clumps 2B, 1D, and 1I, we are mainly observing cold gas (T$_{\rm K}$ about 20 K if we 
assume that the gas is coupled to the dust) with n$_{\rm H_{2}}$ about a few 10$^{3}$ cm$^{-3}$. The analysis yields that clump 2B is 
the densest one among the three analyzed clumps. It is probable that the \2 J=3--2 line arises from both, a cold component and a warmer one.

(5) The maps of the integrated line \2/\3 ratios show  quite uniform values at the peaks of the clumps, with larger
values at the borders, while a gradient is found extending from the border exposed to the N11 central cluster 
to the opposite one in N11D. Selective photodissociation of the \3 can explain what is observed 
in N11B and N11I, while chemical fractionation probably should be taken into account in N11D. 

(6) We obtained an average value of 0.65 in the \2 J=3--2/1--0 ratio among the analyzed molecular clumps,
confirming the assumption done by \citet{galametz16} to estimate the contribution of the $^{12}$CO J=3--2 line 
to the 870 $\mu$m flux. However, the variations among the clumps shows that the direct observation of this line 
could improve the estimation of the actual contribution of the dust emission at 870 $\mu$m.
Maps of the \2 J=3--2 line contribution to the continuum emission at 870 $\mu$m were done.
It was found that the N11I presents the larger line contribution to the 870 $\mu$m continuum among the three studied
regions.

(7) It was found that N11B and N11D present \2 J=3--2/1--0 ratios higher and lesser than the unity, respectively. This
is an evidence that the molecular gas in both regions is affected in differnt ways by the radiation. 
Values in the \2 J=3--2/2--1 ratios across the whole region were found to be similar to the value (about 0.8) measured in typical 
galactic disk clouds. The high \2 J=3--2/2--1 ratios in the northern part of N11D shows that there is stellar feedback from the
LH13 cluster. The same is observed in N11B around the LH10 cluster

\begin{acknowledgements}

We thank the anonymous referee for her/his very helpful comments and corrections.
We wish to thank to M. Galametz for kindly provide us with the LABOCA data.
The ASTE project is led by Nobeyama Radio Observatory (NRO), a branch
of National Astronomical Observatory of Japan (NAOJ), in collaboration
with University of Chile, and Japanese institutes including University of
Tokyo, Nagoya University, Osaka Prefecture University, Ibaraki University,
Hokkaido University, and the Joetsu University of Education.
M.C.P. is a doctoral fellow of CONICET, Argentina. S.P. and M.E.O. are members of the {\sl Carrera del
investigador cient\'\i fico} of CONICET, Argentina.
This work was partially supported by grants awarded by ANPCYT and UBA (UBACyT) from Argentina.
M.R. wishes to acknowledge support from Universidad de Chile VID grant ENL22/18 and partial 
support from CONICYT project Basal AFB-170002.
M.R. and M.C.P. acknowledge support from CONICYT (CHILE) through FONDECYT gran N$^{\rm o}$1140839.

\end{acknowledgements}

%
%

\bibliographystyle{aa} 
\bibliography{citas.bib}

\begin{thebibliography}{42}
\expandafter\ifx\csname natexlab\endcsname\relax\def\natexlab#1{#1}\fi

\bibitem[{{Barb{\'a}} {et~al.}(2003){Barb{\'a}}, {Rubio}, {Roth}, \&
  {Garc{\'{\i}}a}}]{barba03}
{Barb{\'a}}, R.~H., {Rubio}, M., {Roth}, M.~R., \& {Garc{\'{\i}}a}, J. 2003,
  \aj, 125, 1940

\bibitem[{{Deharveng} {et~al.}(2005){Deharveng}, {Zavagno}, \&
  {Caplan}}]{deha05}
{Deharveng}, L., {Zavagno}, A., \& {Caplan}, J. 2005, \aap, 433, 565

\bibitem[{{Deharveng} {et~al.}(2009){Deharveng}, {Zavagno}, {Schuller},
  {Caplan}, {Pomar{\`e}s}, \& {De Breuck}}]{deha09}
{Deharveng}, L., {Zavagno}, A., {Schuller}, F., {et~al.} 2009, \aap, 496, 177

\bibitem[{{Drabek} {et~al.}(2012){Drabek}, {Hatchell}, {Friberg}, {Richer},
  {Graves}, {Buckle}, {Nutter}, {Johnstone}, \& {Di Francesco}}]{drabek12}
{Drabek}, E., {Hatchell}, J., {Friberg}, P., {et~al.} 2012, \mnras, 426, 23

\bibitem[{{Duronea} {et~al.}(2017){Duronea}, {Cappa}, {Bronfman}, {Borissova},
  {Gromadzki}, \& {Kuhn}}]{duro17}
{Duronea}, N.~U., {Cappa}, C.~E., {Bronfman}, L., {et~al.} 2017, \aap, 606, A8

\bibitem[{{Elmegreen} \& {Lada}(1977)}]{elmegreen77}
{Elmegreen}, B.~G. \& {Lada}, C.~J. 1977, \apj, 214, 725

\bibitem[{{Enokiya} {et~al.}(2018){Enokiya}, {Sano}, {Hayashi}, {Tachihara},
  {Torii}, {Yamamoto}, {Hattori}, {Hasegawa}, {Ohama}, {Kimura}, {Ogawa}, \&
  {Fukui}}]{enokiya18}
{Enokiya}, R., {Sano}, H., {Hayashi}, K., {et~al.} 2018, \pasj, 70, S49

\bibitem[{{Evans} {et~al.}(2007){Evans}, {Lennon}, {Smartt}, \&
  {Trundle}}]{evans07}
{Evans}, C.~J., {Lennon}, D.~J., {Smartt}, S.~J., \& {Trundle}, C. 2007, \aap,
  464, 289

\bibitem[{{Federman} {et~al.}(2003){Federman}, {Lambert}, {Sheffer},
  {Cardelli}, {Andersson}, {van Dishoeck}, \& {Zsarg{\'o}}}]{feder03}
{Federman}, S.~R., {Lambert}, D.~L., {Sheffer}, Y., {et~al.} 2003, \apj, 591,
  986

\bibitem[{{Galametz} {et~al.}(2016){Galametz}, {Hony}, {Albrecht}, {Galliano},
  {Cormier}, {Lebouteiller}, {Lee}, {Madden}, {Bolatto}, {Bot}, {Hughes},
  {Israel}, {Meixner}, {Oliviera}, {Paradis}, {Pellegrini}, {Roman-Duval},
  {Rubio}, {Sewi{\l}o}, {Fukui}, {Kawamura}, \& {Onishi}}]{galametz16}
{Galametz}, M., {Hony}, S., {Albrecht}, M., {et~al.} 2016, \mnras, 456, 1767

\bibitem[{Garay {et~al.}(2002)Garay, Johansson, Nyman, Booth, Israel, Kutner,
  Lequeux, \& Rubio}]{garay2002}
Garay, G., Johansson, L., Nyman, L.-{\AA}., {et~al.} 2002, \aap, 389, 977

\bibitem[{{Hatano} {et~al.}(2006){Hatano}, {Kadowaki}, {Nakajima}, {Tamura},
  {Nagata}, {Sugitani}, {Tanab{\'e}}, {Kato}, {Kurita}, {Nishiyama}, {Baba},
  {Ishihara}, \& {Sato}}]{hatano06}
{Hatano}, H., {Kadowaki}, R., {Nakajima}, Y., {et~al.} 2006, \aj, 132, 2653

\bibitem[{{Henize}(1956)}]{henize56}
{Henize}, K.~G. 1956, \apjs, 2, 315

\bibitem[{{Herrera} {et~al.}(2013){Herrera}, {Rubio}, {Bolatto}, {Boulanger},
  {Israel}, \& {Rantakyr{\"o}}}]{herrera2013}
{Herrera}, C.~N., {Rubio}, M., {Bolatto}, A.~D., {et~al.} 2013, \aap, 554, A91

\bibitem[{{Israel} {et~al.}(2003){Israel}, {de Graauw}, {Johansson}, {Booth},
  {Boulanger}, {Garay}, {Kutner}, {Lequeux}, {Nyman}, \& {Rubio}}]{israel03}
{Israel}, F.~P., {de Graauw}, T., {Johansson}, L.~E.~B., {et~al.} 2003, \aap,
  401, 99

\bibitem[{{Israel} \& {Maloney}(2011)}]{israel11}
{Israel}, F.~P. \& {Maloney}, P.~R. 2011, \aap, 531, A19

\bibitem[{{Jameson} {et~al.}(2016){Jameson}, {Bolatto}, {Leroy}, {Meixner},
  {Roman-Duval}, {Gordon}, {Hughes}, {Israel}, {Rubio}, {Indebetouw}, {Madden},
  {Bot}, {Hony}, {Cormier}, {Pellegrini}, {Galametz}, \&
  {Sonneborn}}]{jameson2016}
{Jameson}, K., {Bolatto}, A., {Leroy}, A., {et~al.} 2016, \apj, 825, 12

\bibitem[{{Kawamura} {et~al.}(2009){Kawamura}, {Mizuno}, {Minamidani},
  {Filipovi{\'c}}, {Staveley-Smith}, {Kim}, {Mizuno}, {Onishi}, {Mizuno}, \&
  {Fukui}}]{kawamura09}
{Kawamura}, A., {Mizuno}, Y., {Minamidani}, T., {et~al.} 2009, \apjs, 184, 1

\bibitem[{Keller \& Wood(2006)}]{keller2006}
Keller, S.~C. \& Wood, P.~R. 2006, \apj, 642, 834

\bibitem[{{Kudo} {et~al.}(2011){Kudo}, {Torii}, {Machida}, {Davis}, {Tsutsumi},
  {Fujishita}, {Moribe}, {Yamamoto}, {Okuda}, {Kawamura}, {Mizuno}, {Onishi},
  {Maezawa}, {Mizuno}, {Tanaka}, {Yamaguchi}, {Ezawa}, {Takahashi}, {Nozawa},
  {Matsumoto}, \& {Fukui}}]{kudo11}
{Kudo}, N., {Torii}, K., {Machida}, M., {et~al.} 2011, \pasj, 63, 171

\bibitem[{{Langer} \& {Penzias}(1993)}]{langer93}
{Langer}, W.~D. \& {Penzias}, A.~A. 1993, \apj, 408, 539

\bibitem[{Lucke \& Hodge(1970)}]{lucke1970}
Lucke, P. \& Hodge, P. 1970, The Astronomical Journal, 75, 171

\bibitem[{{MacLaren} {et~al.}(1988){MacLaren}, {Richardson}, \&
  {Wolfendale}}]{mac88}
{MacLaren}, I., {Richardson}, K.~M., \& {Wolfendale}, A.~W. 1988, \apj, 333,
  821

\bibitem[{{Massi} {et~al.}(1997){Massi}, {Brand}, \& {Felli}}]{massi97}
{Massi}, F., {Brand}, J., \& {Felli}, M. 1997, \aap, 320, 972

\bibitem[{{Minamidani} {et~al.}(2008){Minamidani}, {Mizuno}, {Mizuno},
  {Kawamura}, {Onishi}, {Hasegawa}, {Tatematsu}, {Ikeda}, {Moriguchi},
  {Yamaguchi}, {Ott}, {Wong}, {Muller}, {Pineda}, {Hughes}, {Staveley-Smith},
  {Klein}, {Mizuno}, {Nikoli{\'c}}, {Booth}, {Heikkil{\"a}}, {Nyman}, {Lerner},
  {Garay}, {Kim}, {Fujishita}, {Kawase}, {Rubio}, \& {Fukui}}]{minami08}
{Minamidani}, T., {Mizuno}, N., {Mizuno}, Y., {et~al.} 2008, \apjs, 175, 485

\bibitem[{{Mokiem} {et~al.}(2007){Mokiem}, {de Koter}, {Evans}, {Puls},
  {Smartt}, {Crowther}, {Herrero}, {Langer}, {Lennon}, {Najarro}, {Villamariz},
  \& {Vink}}]{mokiem07}
{Mokiem}, M.~R., {de Koter}, A., {Evans}, C.~J., {et~al.} 2007, \aap, 465, 1003

\bibitem[{{Ochsendorf} {et~al.}(2017){Ochsendorf}, {Zinnecker}, {Nayak},
  {Bally}, {Meixner}, {Jones}, {Indebetouw}, \& {Rahman}}]{ochsendorf2017}
{Ochsendorf}, B., {Zinnecker}, H., {Nayak}, O., {et~al.} 2017, Nature
  Astronomy, 1, 784

\bibitem[{Parker {et~al.}(1992)Parker, Garmany, Massey, \&
  Walborn}]{parker1992}
Parker, J.~W., Garmany, C.~D., Massey, P., \& Walborn, N.~R. 1992, \apj, 103,
  1205

\bibitem[{Paron {et~al.}(2014)Paron, Ortega, Cunningham, Jones, Rubio,
  Fari{\~n}a, \& Komugi}]{paron2014}
Paron, S., Ortega, M.~E., Cunningham, M., {et~al.} 2014, \aap, 572, A56

\bibitem[{Paron {et~al.}(2015)Paron, Ortega, Fari{\~n}a, Cunningham, Jones, \&
  Rubio}]{paron2015}
Paron, S., Ortega, M.~E., Fari{\~n}a, C., {et~al.} 2015, \mnras, 455, 518

\bibitem[{{Pomar{\`e}s} {et~al.}(2009){Pomar{\`e}s}, {Zavagno}, {Deharveng},
  {Cunningham}, {Jones}, {Kurtz}, {Russeil}, {Caplan}, \&
  {Comer{\'o}n}}]{pomares09}
{Pomar{\`e}s}, M., {Zavagno}, A., {Deharveng}, L., {et~al.} 2009, \aap, 494,
  987

\bibitem[{{Rathborne} {et~al.}(2002){Rathborne}, {Burton}, {Brooks}, {Cohen},
  {Ashley}, \& {Storey}}]{rath02}
{Rathborne}, J.~M., {Burton}, M.~G., {Brooks}, K.~J., {et~al.} 2002, \mnras,
  331, 85

\bibitem[{{Rosado} {et~al.}(1996){Rosado}, {Laval}, {Le Coarer}, {Georgelin},
  {Amram}, {Marcelin}, {Goldes}, \& {Gach}}]{rosado96}
{Rosado}, M., {Laval}, A., {Le Coarer}, E., {et~al.} 1996, \aap, 308, 588

\bibitem[{{Sawada} {et~al.}(2008){Sawada}, {Ikeda}, {Sunada}, {Kuno},
  {Kamazaki}, {Morita}, {Kurono}, {Koura}, {Abe}, {Kawase}, {Maekawa},
  {Horigome}, \& {Yanagisawa}}]{sawada2008}
{Sawada}, T., {Ikeda}, N., {Sunada}, K., {et~al.} 2008, \pasj, 60, 445

\bibitem[{{Sz{\H u}cs} {et~al.}(2014){Sz{\H u}cs}, {Glover}, \&
  {Klessen}}]{sz14}
{Sz{\H u}cs}, L., {Glover}, S.~C.~O., \& {Klessen}, R.~S. 2014, \mnras, 445,
  4055

\bibitem[{Van~der Tak {et~al.}(2007)Van~der Tak, Black, Sch{\"o}ier, Jansen, \&
  van Dishoeck}]{tak2007}
Van~der Tak, F., Black, J.~H., Sch{\"o}ier, F., Jansen, D., \& van Dishoeck,
  E.~F. 2007, \aap, 468, 627

\bibitem[{{van Dishoeck} \& {Black}(1988)}]{van88}
{van Dishoeck}, E.~F. \& {Black}, J.~H. 1988, \apj, 334, 771

\bibitem[{{Visser} {et~al.}(2009){Visser}, {van Dishoeck}, \&
  {Black}}]{visser09}
{Visser}, R., {van Dishoeck}, E.~F., \& {Black}, J.~H. 2009, \aap, 503, 323

\bibitem[{{Walborn} {et~al.}(1999){Walborn}, {Drissen}, {Parker}, {Saha},
  {MacKenty}, \& {White}}]{walb99}
{Walborn}, N.~R., {Drissen}, L., {Parker}, J.~W., {et~al.} 1999, \aj, 118, 1684

\bibitem[{{Walborn} \& {Parker}(1992)}]{walb92}
{Walborn}, N.~R. \& {Parker}, J.~W. 1992, \apjl, 399, L87

\bibitem[{Wang {et~al.}(2009)Wang, Chin, Henkel, Whiteoak, \&
  Cunningham}]{wang2009}
Wang, M., Chin, Y.-N., Henkel, C., Whiteoak, J., \& Cunningham, M. 2009, \apj,
  690, 580

\bibitem[{{Zavagno} {et~al.}(2010){Zavagno}, {Anderson}, {Russeil}, \& {et
  al.}}]{zav10}
{Zavagno}, A., {Anderson}, L.~D., {Russeil}, D., \& {et al.} 2010, \aap, 518,
  L101

\end{thebibliography}

\end{document}